\documentclass[aps,prd]{revtex4}
\usepackage{epsfig,epsf}
\usepackage{amsmath}
\usepackage{amsthm}
\usepackage{amsfonts}
\usepackage{amssymb}
\usepackage{dsfont}
\usepackage{multirow}
\usepackage{appendix}
\usepackage{slashed}
\usepackage[active]{srcltx}
\usepackage{psfrag}
\usepackage{subfigure}
\usepackage[colorlinks,citecolor=blue,urlcolor=red,linkcolor=red]{hyperref}
\usepackage[colorlinks,citecolor=blue,urlcolor=red,linkcolor=red]{hyperref}
\begin{document}

\title{{\Large{\bf Helicity form factors for $D_{(s)} \to A \ell \nu$ process
in the light-cone QCD sum rules approach}}}

\author{\small
S. Momeni \footnote {e-mail: samira.momeni@ph.iut.ac.ir }}

\affiliation{\emph{ Department of Physics, Isfahan University of
Technology, Isfahan 84156-83111, Iran} }

\begin{abstract}
The helicity form factors of the
$D_{(s)}\to A \ell^+ \nu$ with $A=a_{1}^{-}, a_{1}^{0}, b_{1}^{-}, b_{1}^{0}, K_{1}(1270)$ and $K_{1}(1400)$
 are calculated in the light-cone sum rules approach, up to twist-3 distribution amplitudes of the axial vector
meson $A$. In the helicity form factors parametrization the unitarity constraints are applied
to the fitting parameters. In addition, the effects of the low-lying resonances are included in series expansions of aforementioned
form factors. The properties of
the $D_{(s)}\to A \ell^+ \nu$ semileptonic decays are studied by extending the  form factors to the whole
physical region of $q^2$. For a better analysis, a comparison is also made  between
our results and  the predictions  obtained using transition form factors via LCSR, 3PSR and CLFQM
methods.
\end{abstract}

\pacs{11.55.Hx, 13.20.-v, 14.40.Lb}

\maketitle

\section{Introduction}

The weak semileptonic and hadronic decays of charmed mesons, which occur  in the presence of strong interaction,
 are ideal laboratory candidates  to  determine the
 quark mixing parameters and the values of the Cabibbo-Kobayashi-Maskawa (CKM) matrix elements  and
 establish new physics beyond the standard model (SM).
These meson category masses  are $(\mathcal {O}~ 2 ~\rm{GeV})$, therefore  charm decays are helpful to study
nonperturbative QCD while,  the heavy quark effective
theory (HQET) can  also be utilized to study $D$ meson decays \cite{Artuso}.

$D_{(s)}$ meson decays can be classified into  two categories. The first one, which occurs via $c \to u~ \ell^+ \ell^-$ transition at quark level, is named the flavor changing neutral currents (FCNC) decay. The  $D \to \pi \ell^+
\ell^-$, $D \to \rho \ell^+ \ell^-$, $D \to \pi \,\gamma$ and $D \to
\rho\, \gamma $  from the first group, are studied  using QCD  factorization \cite{Feldmann}.
The second class, which happens by the semileptonic decay of charm quark $c
\to d(s)\ell \nu$ are analyzed   via different  approaches.
Traditionally, semileptonic decays are explained in terms of transition form factors  as a function of
the invariant mass of the electron-neutrino pair, $q^2$. These form factors
which parameterize nonperturbative effects, are measured for $D\to K \ell \nu$ decay
in \cite{Zhang}, while, the Light Cone QCD Sum Rule (LCSR) approach
is utilized to studying $D\to \pi(K, \rho)\,\ell\, \nu$ decays \cite{Khodjamirian, BallD,Fu}.
The form factors of the
 $D^+ \to (D^0, \rho^0, \omega, \eta,
\eta')\ell^+ \nu$ and $D^+_s \to (D^0, \phi, K^0, K^{*0}, \eta,
\eta') \ell^+ \nu $ semileptonic decays have been calculated in the framework of the
covariant confined quark model (CCQM) \cite{Soni1,Soni2}.
The semileptonic
decays $D\to (\pi, \rho, K, K^{*})\ell \nu $  have been studied  using
the  (HQET) in Ref. \cite{WangD} and  the lattice QCD (LQCD) results for
the $D\to
\pi (K, K^{*})\ell\,\nu$ processes  are reported in \cite{Abada, Aubin, Bernard}.
In Ref. \cite{Ignacio,Aliev, Ball2,Ball3,Ovchinnikov,Dong,Mao}
the semileptonic decays
$D_{(s)} \to f_0 (K_0^*)\,\ell\, \nu$, $D_{(s)} \to \pi (K) \,\ell\,
\nu$, and $D_{(s)} \to K^* (\rho, \phi) \,\ell\, \nu$  have been
investigated in the framework of the three-point QCD sum rules (3PSR).
The $D\to\,a_{1},f_{1}(1285), f_{1}(1420)$ and $D_{(s)}\to K_{1}\,\ell\,\nu$
 transitions as the
$D_{(s)}$ decay to the axial vector mesons, have been
calculated by the 3PSR method  \cite{Khosravi,Zuo}.

In this paper, the helicity form factors for the $D_{(s)}$ decays into
axial vectors are  calculated  with the LCSR. The helicity form factors which can be obtained
by contracting the $W (\rm{or}~ Z )$ boson polarization vectors and the
transition matrix elements,  are also
functions of $q^2$. The relations among the $D_{(s)} \to A$ transition
 matrix elements, transition form factors
and the helicity ones are presented in Table \ref{tab:formfact}.

\begin{table}[htb]
\begin{center}
\begin{tabular}{ccl}
\hline
~~Matrix element~~ & ~~Transition form factors~~ & ~~~~~~Helicity form factors~~ \\
\hline
$\begin{array}{c}\langle A|\bar{s}(\bar{d}, \bar{u} )~\gamma^\mu \gamma^{5} c|D_{(s)}\rangle
\\ \langle A|\bar{s}(\bar{d}, \bar{u})~\gamma^\mu c|D_{(s)}\rangle\end{array}$ &
$\begin{array}{c}  A\\ V_0, V_1, V_2 \end{array}$ &
$\left\}\begin{array}{c} \mathcal{H}_{\mathcal{V}, 0} \\
\mathcal{H}_{\mathcal{V}, 1},~\mathcal{H}_{\mathcal{V}, 2}\end{array}\right.$ \\[5mm]
$\begin{array}{c}\langle A|\bar{u} ~\sigma^{\mu\nu}\gamma^5 q_\nu c|D_{(s)}\rangle \\
\langle A|\bar{u} ~\sigma^{\mu\nu} q_\nu c|D_{(s)}\rangle\end{array}$ &
$\begin{array}{c}T_1(q^2)\\T_2(q^2), T_3(q^2)\end{array}$ &
$\left\}\begin{array}{c}\mathcal{H}_{\mathcal{T},0} \\
\mathcal{H}_{\mathcal{T},1},~\mathcal{H}_{\mathcal{T},2}\end{array}\right.$ \\ \hline
\end{tabular}
\caption{ The $D_{(s)}\to A$ decay hadronic matrix elements with the
corresponding transition and  helicity form factors. In this table $\mathcal{V}$ and $\mathcal{T}$ stands for the vector and tensor current, respectively.}
\label{tab:formfact}
\end{center}
\end{table}
There are some advantages in using the  helicity form factors:

1) Diagonalizable unitarity relations can  be imposed on the coefficients of  the helicity form factor
 parameterization.

2) In the helicity form factors, the contributions from
the excited states  and the spin-parity
quantum numbers are considered by relating the dominant poles in the
LCSR predictions to low-lying resonances (for more detailed, see \cite{Bharucha2010}).

The masses and quantum numbers $J^{P}$ of low-lying $D_{(s)}$ resonances with the relations among the
helicity form factors
are provided in Table \ref{Tm}. These masses will be used in the helicity form factors parameterizations.
Notice that the mass values for  $D_{s} (1^{-})$ and none of the  $(1^{+})$ states predicted in \cite{Bardeen2003} have  been experimentally confirmed  yet.
\begin{table}[htb]
\begin{center}
\begin{tabular}{|cccc|ccccc|}
\hline
&&~~$\rm{D\, meson}$~~ & ~~$\rm{Mass (Gev)}$~~  && &&~~$\rm{D \,meson}$~~ & ~~$\rm{Mass (Gev)}$~~ \\
\hline
$\begin{array}{c}  J^{P}=1^{-}, \\
\mathcal{H}_{\mathcal{V}, 1}\\
\end{array}$&
$\left\{\begin{array}{c}  \\
\\
\end{array}\right.$&
$\begin{array}{c} D^{+}
\\ D^{0}
\\D_{s}
\end{array}$ &
$\begin{array}{c} 2.01
\\ 2.00
\\2.11
\end{array}$&
$\begin{array}{c}  J^{P}=1^{+}, \\
\mathcal{H}_{\mathcal{V}, 0}, \mathcal{H}_{\mathcal{V}, 2}\\
\end{array}$&
$\left\{\begin{array}{c}  \\
\\
\end{array}\right.$&
$\begin{array}{c}  \\
\\
\end{array}$&
$\begin{array}{c} D^{+}
\\ D^{0}
\\D_{s}
\end{array}$ &
$\begin{array}{c} 2.35
\\ 2.35
\\2.46
\end{array}$\\
\hline
\end{tabular}
\caption{ The masses of low-lying $D_{(s)}$ resonances and their relations to the helicity form factors. The masses
 are taken form PDG values
\cite{pdg} and the heavy-quark chiral symmetry approach \cite{Bardeen2003}.}
\label{Tm}
\end{center}
\end{table}

In \cite{Cheng2018},
the helicity   form factors are
calculated via LCSR approach for $B\to \rho$ decay. In this paper, these  form factors are evaluated for
$D^{0} \to a^{-}_{1} (b^{-}_{1}) \ell^+ \nu$, $D^{+}
\to a^{0}_{1} (b^{0}_{1}) \ell^+ \nu$ and $D_{(s)}\to K_{1} \ell^+ \nu$ decays, which are described by $c \to d~ \ell
\nu$ transition at quark level. The form factors  are also estimated for the   $c
\to s~ \ell \nu$ transition in
$D^{+} \to K_{1}
\ell^+ \nu$ decay. Here the physical states $K_{1}=K_{1} (1270), K_{1} (1400)$
are the mixtures of the $K_{1A}$ and $K_{1B}$  in terms of a mixing angle as \cite{Kwei2}:
\begin{eqnarray}\label{eq21}
|K_1(1270)\rangle &=&\sin\theta_K |K_{1A}\rangle +
\cos\theta_K |K_{1B}\rangle,\nonumber\\
|K_1(1400)\rangle &=&\cos\theta_K |K_{1A}\rangle - \sin\theta_K
|K_{1B}\rangle,
\end{eqnarray}
where $|K_{1A}\rangle$ and $|K_{1B}\rangle$ are not mass eigenstates. The mixing angle $\theta_K$  is determined by
various experimental analyses.  The result $35^\circ \leq |\theta_K| \leq
55^\circ$ was reported in Ref. \cite{Burakovsky}. Moreover, two possible
solutions were obtained as $|\theta_K|\approx 33^\circ \vee
57^\circ$  in Ref. \cite{Suzuki} and as $|\theta_K|\approx 37^\circ
\vee 58^\circ$ in Ref. \cite{HYCheng}. Using the study of
 $B\to K_1(1270)\gamma$
and $\tau \to  K_1(1270)\nu_{\tau}$ decays, the value of
$\theta_K$ is estimated as \cite{Hatanaka2}
\begin{eqnarray}
\theta_K = {-(34\pm13)}^{\circ}.
\end{eqnarray}
 In this study,  the branching ratio values are reported for  the  $D_{(s)} \to K_{1}
\ell^+ \nu$ decays at  $\theta_K = {-(34\pm13)}^{\circ}$.

Our paper is organized as follows: In Sec. \ref{se.c}
by using the
LCSR, the form factors of $D_{(s)} \to A$ decays are derived.
Section. \ref{se.nu}, is devoted to the numerical analysis of the form factors
and the branching ratios  for  semileptonic and  decays.
A comparison of our results for the branching ratios
 with the other
approaches and existing experimental data is also made in this section and the last section is reserved for summary.

\section{Light cone QCD sum rules for $D^{0} \to a_{1}^{-} \ell \nu$ Helicity form factors }\label{se.c}
To calculate the helicity form factors of $D^{0} \to a_{1}^{-} \ell \nu$ decay,
the following correlation function is considered:
\begin{eqnarray}\label{eq.1}
\Pi_{\sigma}^{a_{1}} (p_{i}, p_{f})&=& \sqrt{\frac{q^2}{\lambda}}\sum_{\alpha}\varepsilon^{*\mu}_{\sigma}\int d^4x\,
 e^{iqx} \langle a_{1}^{-} (p_{f}, \varepsilon_{\alpha}) | {\cal{T}} \{\,j_{\mu}^{int}(x)\,j_{D^0}^{\dag}(0) \}| 0 \rangle,
\end{eqnarray}
where $p_{i}$,  $p_{f}=(p_{f}^0,0,0,|\vec{p_{f}}|)$ and  $q=p_{i}-p_{f}$ are the
four-momentum of the  $D^{0}$, $a_{1}^{-}$  and $W$-boson, respectively.
Moreover,  $j_{\mu}^{int}=\bar{d}\gamma_\mu (1-\gamma_5) c$ is
the interaction current for $D^{0} \to a_{1}^{-} $
process and $j_{D^0}=i{\bar u}\,\gamma_5\,c$ is the  interpolating current
for $D^0$ meson.
In $\Pi_{\sigma}^{a_{1}}$ expression, $\varepsilon_{\alpha}$ and $\varepsilon_{\sigma}$
denote the polarization for $a_{1}$ meson and $W$-boson, respectively as
\begin{eqnarray}\label{eq.2}
\varepsilon _{{\alpha=0}} &=&  \frac{1}{m_{a_{1}^{-}}} (|\vec{p_{f}}| ,0,0,p_{f}^0),  \\
\varepsilon _{{\alpha=\pm}} &=&  \mp \frac{1}{{\sqrt 2 }}  (0,1, \mp i,0),\\
\varepsilon _{{\sigma=0}} &=&  \frac{1}{{\sqrt {q^2 } }} \, (|\vec q| ,0,0, -q^0 ),
\end{eqnarray}
with $|\vec{p}_{f}|=\sqrt{\lambda}/2m_{D^{0}}$, $p_{f}^0 = {(m_{D^{0}}^2 + m_{a_{1}^{-}} ^2-q^2)}/{2m_{D^{0}}}$,
 $|\vec q|=|\vec {p}_{f}|$ and $q^0 = {(m_{D^{0}}^2 - m_{a_{1}^{-}} ^2+q^2)}/{2m_{D^{0}}}$.
Also, $\lambda = (t_{-} - q^2)(t_{+} - q^2)$ with $t_\pm=(m_{D^{0}}\pm m_{a_{1}^{-}})^2$. Moreover,  $\varepsilon _{{\sigma=\pm}}$
has similar definition as  $\varepsilon _{{\alpha=\pm}}$.

For off-shell $W$-boson,
$\varepsilon _{{\sigma=1}}$ and $\varepsilon _{{\sigma=2}}$ are linear
 combinations of the transverse $(\pm)$ polarization vectors as
\begin{eqnarray}\label{eq.3}
 \varepsilon _{\sigma=1}  &=& \frac{{(\varepsilon _{\sigma=-}) - (\varepsilon _ {\sigma=+}) }}{{\sqrt 2 }} = (0,1,0,0), \\
\varepsilon _{\sigma=2}  &=& \frac{{(\varepsilon _{\sigma=-)} + (\varepsilon _ {\sigma=+})}}{{\sqrt 2 }} = (0,0,i,0).
\end{eqnarray}
In the Light Cone QCD sum rules approach, the correlation function is given in Eq. (\ref{eq.1}), should be calculated  in  phenomenological and theoretical representations. Helicity form factors  are found to equate both representations
of the correlation function through dispersion relation.

The phenomenological side can be obtained by
inserting a complete series of the intermediate hadronic states with the same quantum numbers as the interpolating current
 $j_{D^0}$. After separating the lowest $D^{0}$ meson ground state
 and applying Fourier transformation, $\Pi_{\sigma}^{a_{1}}$ is
obtained as:
\begin{eqnarray}\label{eq.4}
\Pi _{\sigma}^{a_{1} }&=&\sqrt{\frac{q^2}{\lambda}} \, \sum_{\alpha}
{\varepsilon_\sigma^{*\mu}}\frac{\langle a_{1}^{-}(p_{f},\varepsilon_\alpha)
|\bar d \, \gamma_\mu(1-\gamma^5) \, c |D^{0}\rangle \langle D^{0}|j_{D^0}^{\dag}(0)|0\rangle }{(m_{D^{0}}^2 - p_{i}^{2})}
\nonumber \\
&+&\frac{1}{\pi}\sqrt{\frac{q^2}{\lambda}}\sum_{\alpha}{\varepsilon_\sigma^{*\mu}}
\int_{0}^{\infty}\frac{\rho_{\mu}^{h}(s)}{s-p^2}\,ds,
\end{eqnarray}
where, $\rho_{\mu}$ is the density of  higher
states and continuum which can be approximated
using the ansatz of the quark-hadron duality as
\begin{eqnarray}\label{eq.rho}
\rho_{\mu}^{h}(s)=\rho_{\mu}^{QCD}(s)\theta(s-s_{0}),
\end{eqnarray}
where, $\rho_{\mu}^{QCD}=\frac{1}{\pi}\rm{Im}(\Pi _{\mu}^{QCD})$ is the perturbative QCD
spectral density and  $s_{0}$ is the continuum threshold in $D^{0}$ channel.
Now, the following definitions are used for the first and second matrix elements in Eq. (\ref{eq.4}):
\begin{eqnarray}\label{eq.5}
\sqrt{\frac{q^2}{\lambda}} \, \sum_{\alpha}
{\varepsilon_\sigma^{*\mu}}{\langle a_{1}^{-}(p_{f},\varepsilon_\alpha)
|\bar d \, \gamma_\mu(1-\gamma^5) \, c |D^{0}\rangle}=\mathcal{H}_{\sigma}^{a_{1}^{-}},~~
\langle D^{0}|j_{D^0}^{\dag}(0)|0\rangle=\frac{f_{D^{0}}\,m_{D^{0}}^{2}}{m_{c}},
\end{eqnarray}
where $\mathcal{H}_{\sigma}^{a_{1}^{-}}$, $f_{D^{0}}$ and $m_{D^{0}}$  are the helicity form factor of $D^{0} \to a_{1}^{-} \ell \nu$ decay, the decay constant and mass of the ${D^{0}}$ meson, respectively. The final result for phenomenological
part of correlation function is obtained as:
\begin{eqnarray}\label{eq.6}
\Pi _{\sigma}^{a_{1} }&=&\frac{f_{D^{0}}\,m_{D^{0}}^{2}}{m_{c}}
\frac{\mathcal{H}_{\sigma}^{a_{1}^{-}}}{(m_{D^{0}}^2 - p_{i}^{2})}
+\frac{1}{\pi}\sqrt{\frac{q^2}{\lambda}}\sum_{\alpha}{\varepsilon_\sigma^{*\mu}}
\int_{0}^{\infty}\frac{\rho_{\mu}^{h}(s)}{s-p^2}\,ds,
\end{eqnarray}
To evaluate the correlation function $\Pi _{\sigma}^{a_{1} }$ in QCD side, the $\mathcal{T}$ product of currents should be expanded
near the light cone $x^2\simeq0$. After
contracting $c$ quark field,
\begin{eqnarray}\label{eq.th1}
\Pi_{\sigma}^{a_{1}} (p_{i}, p_{f})&=& -i\sqrt{\frac{q^2}{\lambda}}\sum_{\alpha}\varepsilon^{*\mu}_{\sigma}\int d^4x\,
 e^{iqx} \langle a_{1}^{-} (p_{f}, \varepsilon_{\alpha}) | \bar d(x) \gamma_{\mu}\, (1-\gamma_{5})\,S_{c}(x, 0)\,c(0)\}| 0 \rangle,
\end{eqnarray}
is obtained.
Where $S_{c}(x, 0)$ is the full propagator of the $c$ quark. In this paper, the contributions from the gluon contributions
have been neglected  and
 only the free propagator is considered as:
\begin{eqnarray}\label{prop}
S_{c}(x, 0)=\int \frac{d^4l}{(2\pi)^4} e^{-il.x} \frac{\not\!l +
m_c}{l^2-m_c^2}
\end{eqnarray}
Replacing Eq. (\ref{prop}) in theoretical part of $\Pi_{\sigma}^{a_{1}} (p_{i}, p_{f})$ yields:
\begin{eqnarray}\label{eq.th2}
\Pi_{\sigma}^{a_{1}} (p_{i}, p_{f})&=&-i\sqrt{\frac{q^2}{\lambda}}\sum_{\alpha}\int \frac{d^4l}{(2\pi)^4} \int d^4x\,
\frac{e^{i(q-l)x}}{l^2-m_c^2}\nonumber\\
&\times&\Bigg\{\varepsilon^{*\mu}_{\sigma}\,l^\nu \langle a_{1}^{-} (p_{f}, \varepsilon_{\alpha})|  \bar d(x)\gamma _\mu \gamma _\nu \gamma _5 c(0) |0\rangle +\varepsilon^{*\mu}_{\sigma} {l^\nu }\langle a_{1}^{-} (p_{f}, \varepsilon_{\alpha})|\bar d(x)\gamma _\mu \gamma _\nu c(0)\}|0\rangle  \nonumber\\
&& +m_{c}~\varepsilon^{*\mu}_{\sigma} \langle a_{1}^{-} (p_{f}, \varepsilon_{\alpha})| \bar d(x) \gamma_\mu \gamma_5 c(0)\}|0\rangle
-m_{c}~\varepsilon^{*\mu}_{\sigma}\langle a_{1}^{-} (p_{f}, \varepsilon_{\alpha})| \bar d(x) \gamma_\mu c(0) \}|0\rangle . \Bigg\}.
\end{eqnarray}
As it is clear from  Eq. (\ref{eq.th2}),  to
calculate the theoretical part of the correlation  function, the matrix elements
of the nonlocal operators between $a_{1}^{-}$ meson and vacuum state are needed.
Two- particle distribution amplitudes  for the axial vector mesons are given in \cite{Kwei},
which are put in the Appendix.

In this step, two-particle LCDAs are inserted in Eq. (\ref{eq.th2}), and then integrals over $x$ and $l$ should be
evaluated. To estimate
these calculations, the following identities are utilized:
\begin{eqnarray}
\gamma _\mu \gamma _\nu  &=& g_{\mu \nu } - i\sigma _{\mu \nu }, \\
\gamma _\mu \gamma _\nu \gamma _5 &=& g_{\mu \nu }\gamma _5 - \frac{1}{2}\varepsilon _{\mu \nu \rho \beta }\sigma ^{\rho \beta },\\
\gamma _5 \sigma ^{\mu \nu} &=&  - \frac{i}{2}\sigma ^{\rho \beta }\varepsilon _{\mu \nu \rho \beta },\\
\epsilon_{\kappa\nu\beta\mu}\,\epsilon^{\nu\beta\rho\omega}&=&
2\,(\delta^{\rho}_{\kappa}\,\delta^{\omega}_{\mu}-\delta^{\omega}_{\kappa}\,\delta^{\rho}_{\mu}).
\end{eqnarray}
Now, to get the LCSR  calculations for  the $D^{0} \to a_{1}^{-} \ell \nu$
helicity form factors,  the expressions for $\sigma=0, 1, 2 $ from both
phenomenological and theoretical
sides of the correlation function are equated  and  Borel transform is applied
with respect to  variable $p_{i}^{2}$ as:
\begin{eqnarray}\label{eqborel}
B_{p_{i}^2}(M^2)\frac{1}{\left(
p_{i}^{2}-m_{D^0}^{2}\right)^{n}}&=&\frac{(-1)^{n}}{\Gamma(n)}\frac{e^{-\frac{m_{D^0}^{2}}{M^{2}}}}{(M^{2})^{n}},
\end{eqnarray}
which eliminates the subtraction term in
the dispersion relation and exponentially suppresses the
contributions of higher states. Finally, the helicity form factors for $D^{0} \to a_{1}^{-} \ell \nu$, 
transition are obtained in the LCSR as
\begin{eqnarray}
\mathcal{H}_{0}^{a_{1}^{-}}&=&\frac{m_{c}f_{a_{1}^{-}}}{f_{D^{0}}m_{D^{0}}^{2}}\Bigg\{-\frac{m_{c}}{4}\int_{u_0}^{1}du\,
\frac{\left.[\Phi_{\perp}^{i}(u)-g_{\perp}^{i,(a)}\right.]~\sqrt{\lambda}}{u^2\,M^{2}}~e^{\frac{s(u)}{M^{2}}}+
\frac{m_{c}}{\sqrt{\lambda}}\int_{u_0}^{1}du
\frac{g_{\perp}^{i,(a)}~\theta_{1}(q^2)}{u}~e^{\frac{s(u)}{M^{2}}}\nonumber\\
&+&\frac{m_{a_{1}^{-}}}{4}\int_{u_0}^{1}du\,
\frac{h_{\parallel}^{(p)}(u)~(u+1)~\sqrt{\lambda}}{u^2\,M^{2}}~e^{\frac{s(u)}{M^{2}}}+
\frac{~f_{a_{1}^{-}}^{\perp}}{m_{a_{1}^{-}}f_{a_{1}^{-}}}\int_{u_0}^{1}du\,
\frac{\Phi_{\perp}(u)}{u}\left.[\frac{2~\theta_{1}(q^2)~\theta_{2}(q^2, u)+\lambda}{\sqrt{\lambda}}\right.]~e^{\frac{s(u)}{M^{2}}}\nonumber\\
&+&\frac{4~f_{a_{1}^{-}}^{\perp}~m_{a_{1}^{-}}}{f_{a_{1}^{-}}}
\int_{u_0}^{1}du\,
\frac{\bar{h}_{\parallel}^{ii, (t)}(u)~\sqrt{\lambda}}{u^2\,M^{2}}~\left.[1+\frac{~\theta_{3}(q^2,u)-\theta_{2}(q^2,u) }{M^{2}}\right.]~e^{\frac{s(u)}{M^{2}}}+\frac{4~f_{a_{1}^{-}}^{\perp}~m_{a_{1}^{-}}}{f_{a_{1}^{-}}}\nonumber\\
&\times&
\int_{u_0}^{1}du\,
\frac{\bar{h}_{\parallel}^{ii, (t)}(u)}{u^2}~\left.[1-\frac{\theta_{2}(q^2,u) }{M^{2}}\right.]~e^{\frac{s(u)}{M^{2}}}\Bigg\},\label{HFF0}
\\
\mathcal{H}_{1}^{a_{1}^{-}}&=&-\frac{m_{c}f_{a_{1}^{-}}}{f_{D^{0}}m_{D^{0}}^{2}}
\sqrt{\frac{q^2}{2}}\Bigg\{\frac{m_{c}\,m_{a_{1}^{-}}}{2}\int_{u_0}^{1}du\,
\left.[\frac{g_{\perp}^{(a)}}{u~\sqrt{\lambda}}+\frac{g_{\perp}^{(v)}}{u^2~M^2}\right.]~e^{\frac{s(u)}{M^{2}}}
-\frac{8~f_{a_{1}^{-}}^{\perp}}{f_{a_{1}^{-}}}\int_{u_0}^{1}du\,
\frac{\Phi_{\perp}(u)}{u}~e^{\frac{s(u)}{M^{2}}}\nonumber\\
&-&\frac{32~f_{a_{1}^{-}}^{\perp}~m_{a_{1}^{-}}^{2}}{f_{a_{1}^{-}}}\int_{u_0}^{1}du\,
\frac{\bar{h}_{\parallel}^{ii, (t)}(u)}{u^2~M^{2}}~e^{\frac{s(u)}{M^{2}}}-
\frac{4~f_{a_{1}^{-}}^{\perp}}{f_{a_{1}^{-}}}\int_{u_0}^{1}du\,
\frac{\Phi_{\perp}(u)~\theta_{2}(q^2, u)}{u~\sqrt{\lambda}}~e^{\frac{s(u)}{M^{2}}}+\frac{4~f_{a_{1}^{-}}^{\perp}~m_{a_{1}^{-}}^{2}}{f_{a_{1}^{-}}}\nonumber\\
&\times&\int_{u_0}^{1}du\,
\frac{\bar{h}_{\parallel}^{ii, (t)}(u)}{u^2~\sqrt{\lambda}}\left.[1-\frac{\theta_{2}(q^2, u)}{M^{2}}\right.]~e^{\frac{s(u)}{M^{2}}}
\Bigg\},\label{HFF1}\\
\mathcal{H}_{2}^{a_{1}^{-}}&=&\frac{m_{c}f_{a_{1}^{-}}}{f_{D^{0}}m_{D^{0}}^{2}}
\sqrt{\frac{q^2}{2}}\Bigg\{\frac{m_{c}\,m_{a_{1}^{-}}}{2}\int_{u_0}^{1}du\,
\left.[\frac{g_{\perp}^{(a)}}{u~\sqrt{\lambda}}-\frac{g_{\perp}^{(v)}}{u^2~M^2}\right.]~e^{\frac{s(u)}{M^{2}}}
-\frac{8~f_{a_{1}^{-}}^{\perp}}{f_{a_{1}^{-}}}\int_{u_0}^{1}du\,
\frac{\Phi_{\perp}(u)}{u}~e^{\frac{s(u)}{M^{2}}}\nonumber\\
&-&\frac{32~f_{a_{1}^{-}}^{\perp}~m_{a_{1}^{-}}^{2}}{f_{a_{1}^{-}}}\int_{u_0}^{1}du\,
\frac{\bar{h}_{\parallel}^{ii, (t)}(u)}{u^2~M^{2}}~e^{\frac{s(u)}{M^{2}}}+
\frac{4~f_{a_{1}^{-}}^{\perp}}{f_{a_{1}^{-}}}\int_{u_0}^{1}du\,
\frac{\Phi_{\perp}(u)~\theta_{2}(q^2, u)}{u~\sqrt{\lambda}}~e^{\frac{s(u)}{M^{2}}}-\frac{4~f_{a_{1}^{-}}^{\perp}~m_{a_{1}^{-}}^{2}}{f_{a_{1}^{-}}}\nonumber\\
&\times&\int_{u_0}^{1}du\,
\frac{\bar{h}_{\parallel}^{ii, (t)}(u)}{u^2~\sqrt{\lambda}}\left.[1-\frac{\theta_{2}(q^2, u)}{M^{2}}\right.]~e^{\frac{s(u)}{M^{2}}}
\Bigg\},\label{HFF2}
\end{eqnarray}
where, $\Phi_\parallel$, $\Phi_\perp$ are twist-2, $g_\perp^{(a)}$,
$g_\perp^{(v)}$, $h_\parallel^{(t)}$ and $h_\parallel^{(p)}$ are
twist-3  functions and $\bar h_\parallel^{(t)} =h_\parallel^{(t)}- \frac{1}{2}
\Phi_\perp(u)$. Moreover, $f_{a_{1}^{-}}$ and $f_{a_{1}^{-}}^{\perp}$ are  scale-independent  scale-dependent decay constants
of the $a_{1}^{-}$ meson, respectively \cite{Kwei}. We also have:
\begin{eqnarray}
u_{0}(s_0) &=&\frac{1}{2m^2_{a_1^{-}}}
\left[\sqrt{(s_0-m_{a_1^{-}}^2-q^2)^2 +4 m_{a_1^{-}}^2 (m_c^2-q^2)}
-(s_0-m_{a_1^{-}}^2-q^2)\right],\nonumber\\
s(u)&=&-\frac{1}{u}\left[m_c^2+u\,\bar{u}m_{a_1^{-}}^2-\bar{u}q^2-u\,m_{D^{0}}^{2}\right],\nonumber\\
\theta_{1}(q^2)&=&\frac{1}{2}(m_{D^{0}}^2-m_{a_{1}^{-}}^2-m_{a_{1}^{-}}^2\frac{q^2}{m_{D^{0}}^2}),\nonumber\\
\theta_{2}(q^2, u)&=&\frac{1}{u}(3~u^2~m_{a_{1}^{-}}^2+q^2-m_{c}^{2}),\nonumber\\
\theta_{3}(q^2, u)&=&\frac{1}{u}(-2~u\,\bar{u}~m_{a_{1}^{-}}^2+2\bar{u}q^2-m_{c}^{2}),\nonumber\\
{f}^{(i)}(u)&\equiv&\int_0^u f(v) dv,~~~~~~~~~{f}^{(ii)}(u)\equiv\int_0^u dv\int_0^v d\omega~ f(\omega).
\end{eqnarray}
The explicit expressions for twist functions are presented in the Appendix.

Following the previous steps in this section, phrases similar to Eqs.
(\ref{HFF0}, \ref{HFF1}, \ref{HFF2}) can be obtained
for the helicity form factors of $D^{0} \to b^{-}_{1} \ell^+ \nu$,
$D^{+} \to a^{0}_{1} (b^{0}_{1}) \ell^+ \nu$, $D^{0} \to K_{1A} \ell^+ \nu$,  $D^{0}\to K_{1B} \ell^+ \nu$, $D_{s} \to
K_{1A} \ell^+ \nu$ as well as $D_{s} \to
K_{1B} \ell^+ \nu$ decays. For the  physical states $K_{1}(1270)$ and
$K_{1}(1400)$   the following relations are used:
\begin{eqnarray*}
\mathcal{H}_{\sigma}^{K_{1}(1270)}&=&\sin\theta_K~\mathcal{H}_{\sigma}^{K_{1A}}+ \cos\theta_K~\mathcal{H}_{\sigma}^{K_{1B}},\\
\mathcal{H}_{\sigma}^{K_{1}(1400)}&=&\cos\theta_K~\mathcal{H}_{\sigma}^{K_{1A}}- \sin\theta_K~\mathcal{H}_{\sigma}^{K_{1B}}.
\end{eqnarray*}

\section{Numerical analysis}\label{se.nu}
Our numerical analysis for the helicity form factors and branching
ratio values of the semileptonic $D_{(s)} \to A \ell^+ \nu$,  are presented in  two subsections.
The helicity
form factors of the semileptonic $D^{+}
\to a^{0}_{1} (b^{0}_{1}, K_{1A}^{0}, K_{1B}^{0} ) \ell^+ \nu$, $D^{0} \to
a^{-}_{1} (b^{-}_{1} ) \ell^+ \nu$, and $D^{+}_{s}\to K_{1A}^{0} (K_{1B}^{0})
\ell^+ \nu$ decays are evaluated in the first subsection.  In the second ones,
using these form factors, the branching ratio values are estimated for considering decays.

In this work, masses are taken in
GeV as  $\mbox{GeV}$ as $m_c=(1.28\pm
0.03)$, $m_{D^{+}(D^{0})}=1.86  $,  $m_{D_s}= 1.96 $, $m_{a_1}=(1.23\pm0.40)$, $m_{b_{1}}=(1.23\pm0.32)$ \cite{pdg},
$m_{K_{1A}}=(1.31\pm0.06)$ and $m_{K_{1B}}=(1.34\pm0.08)$ \cite{Kwei}. The
results of the QCD sum rules are used for decay constants of $D$ and $D_s$ and axial vector
mesons in $\rm {MeV}$,  as $f_{D^{+}(D^{0})}=(210 \pm {12})$ and $f_{D_s}= (246 \pm 8)$ \cite{Mutuk},
$f_{a_{1}}=(238 \pm 10)$, $f_{b_{1}}=(180 \pm 8)$, $f_{K_{1A}}=(250 \pm 13) $ and
  $f_{K_{1B}}= (190 \pm 10)$  \cite{Kwei}.
  We can take $f_{A}=f^\perp_{A}$ at energy scale
$\mu=1\,\rm{GeV}$ \cite{Kwei}. The values of  Gegenbauer
moment for the axial vector mesons, can be found in \cite{Kwei}.

\subsection{Analysis of helicity form factors }
The formulas of helicity from  factors, Eqs.
(\ref{HFF0}, \ref{HFF1}, \ref{HFF2}),    contain
two free parameters $s_{0}$ and $M^{2}$, which are the continuum threshold and  Borel
mass--square, respectively.
In this paper the values of  continuum threshold
are chosen as $s_{0}=(7 \pm 0.2)~\rm{GeV}^{2}$  \cite{Zuo} and  working region for $M^2$
is provided  that the contribution of higher states as well as
higher twist contributions,  be small.

\begin{figure}
\includegraphics[width=5.5cm,height=5.5cm]{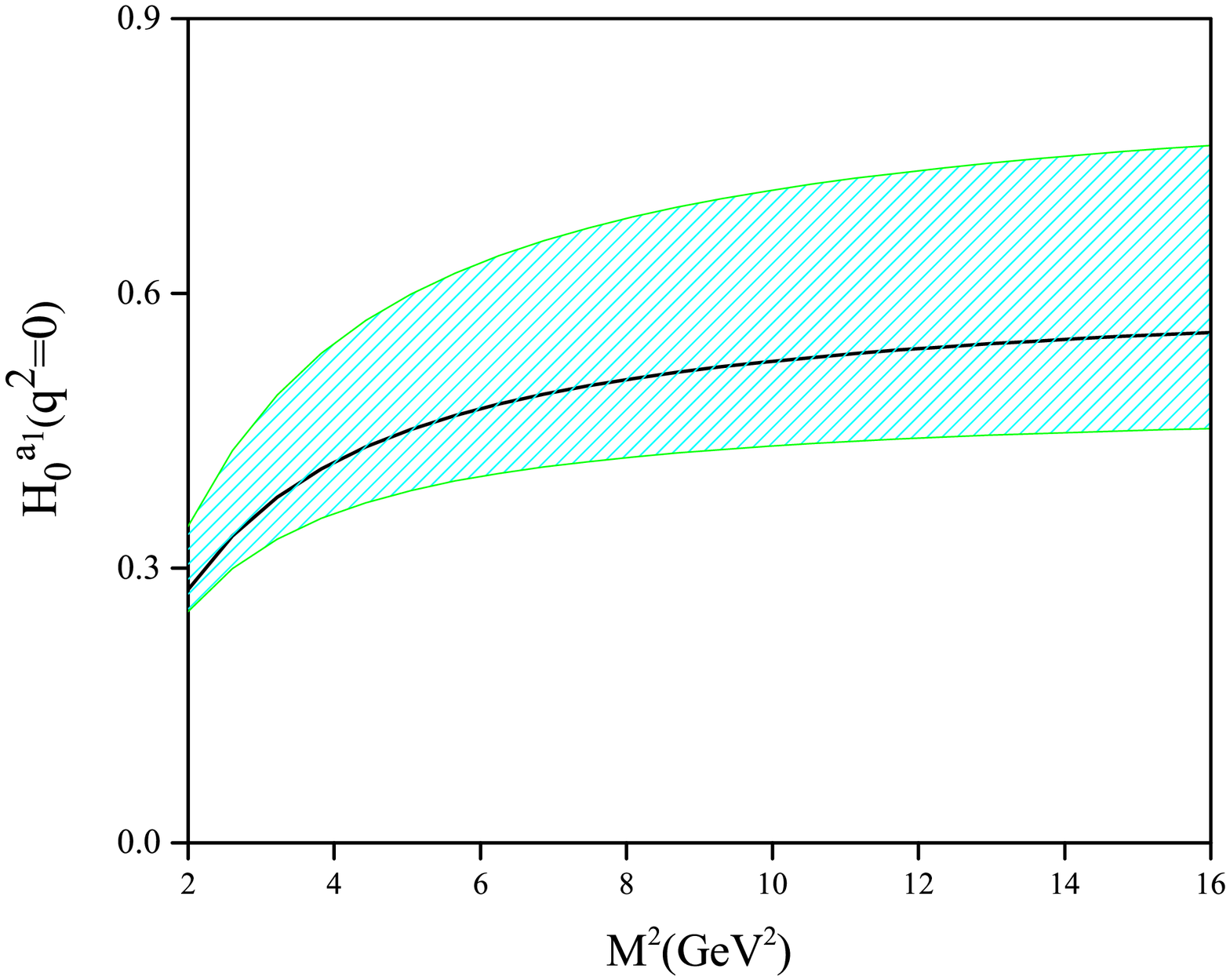}
\includegraphics[width=5.5cm,height=5.5cm]{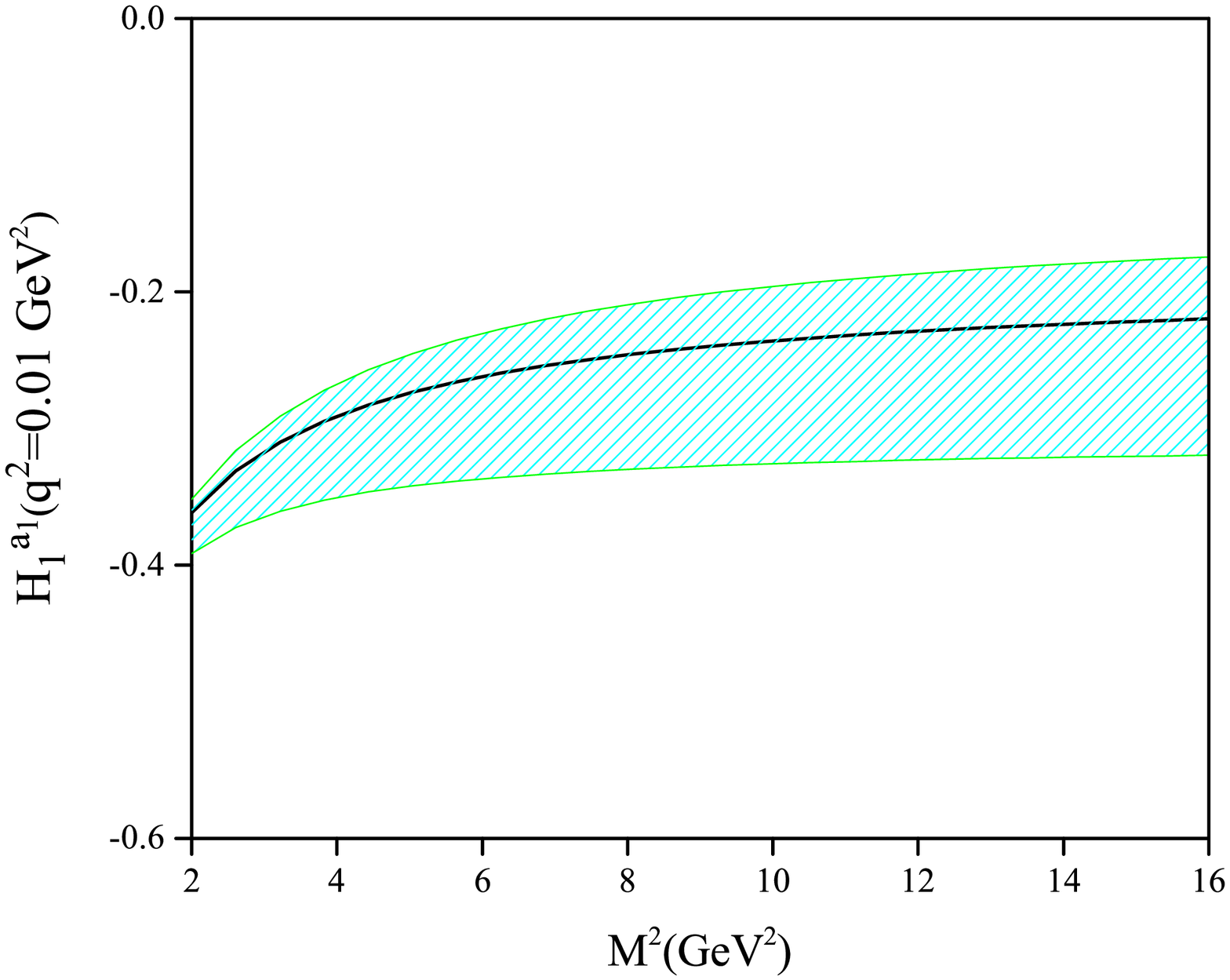}
\includegraphics[width=5.5cm,height=5.5cm]{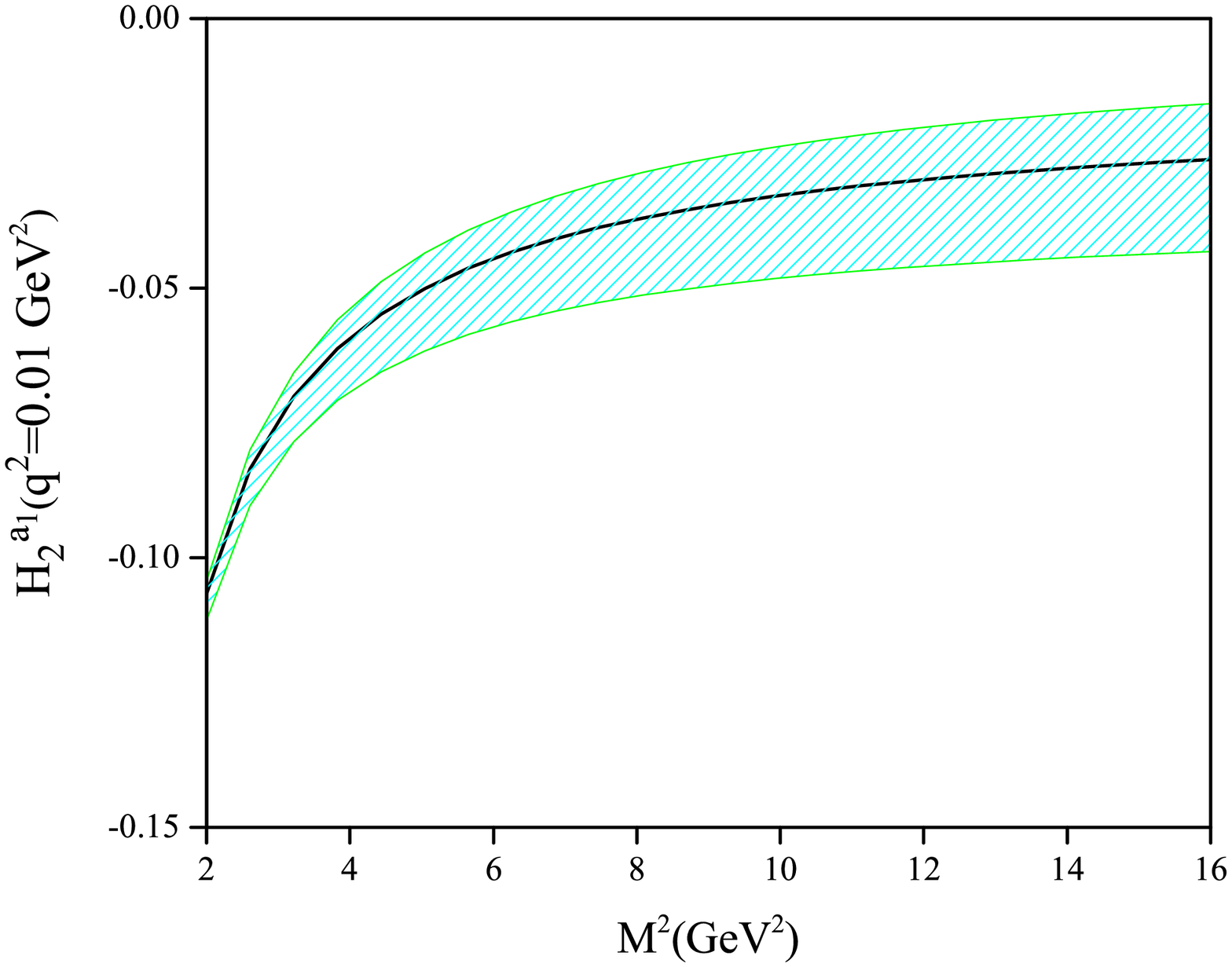}
\caption{$D^0 \to a_1^{-}$ helicity form factors as functions of
$M^{2}$. For $\mathcal{H}_{0}^{a_{1}}  $ we take $q^2=0$ while, for $\mathcal{H}_{1, 2}^{a_{1}}  $
the results are plotted at $q^2=0.01~ \rm{GeV}^{2}$. The threshold parameter is taken $s_{0}=(7 \pm 0.2)~\rm{GeV}^{2}$ for every plot.}\label{Fb}
\end{figure}

Fig. \ref{Fb} shows the dependence of the $D^0 \to a_1^{-}$ helicity form factors  with respect to $M^{2}$. Since  $\mathcal{H}_{\sigma=1, 2}$ vanish at $q^2=0$, these two form factors are plotted at $q^2=0.01~\rm{GeV}^{2}$.
It is easily seen from Fig. \ref{Fb}, that the form factors $\mathcal{H}^{a_{1}^{-}}_{0}$, $\mathcal{H}^{a_{1}^{-}}_{1}$
and $\mathcal{H}^{a_{1}^{-}}_{ 2}$ obtained
from the sum rules,  can be stable within
the Borel parameter intervals  $5~ \mbox{GeV}^2 <M^{2}<8~ \mbox{GeV}^2$.

\begin{figure}[th]
\includegraphics[width=5.5cm,height=5.5cm]{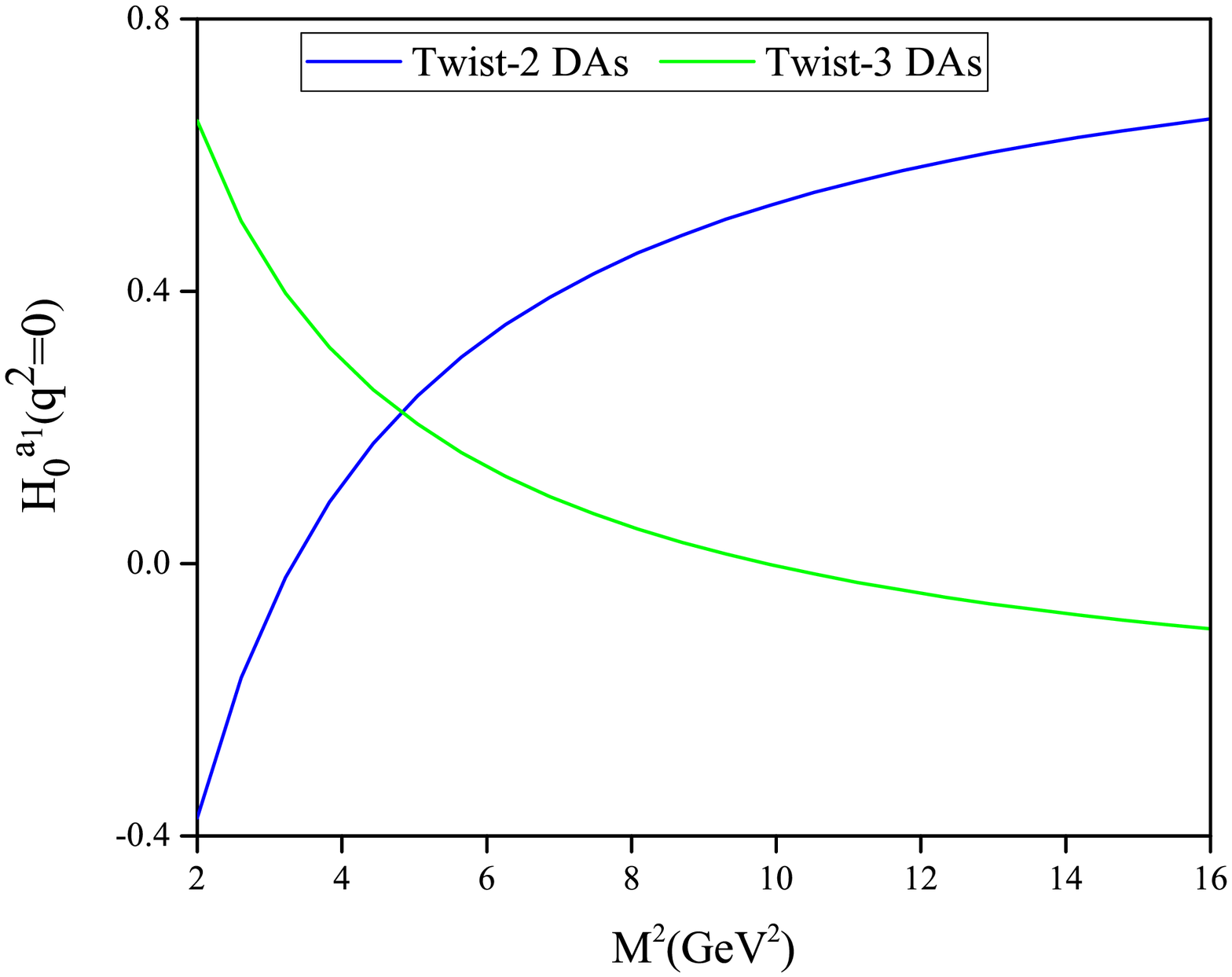}
\includegraphics[width=5.5cm,height=5.5cm]{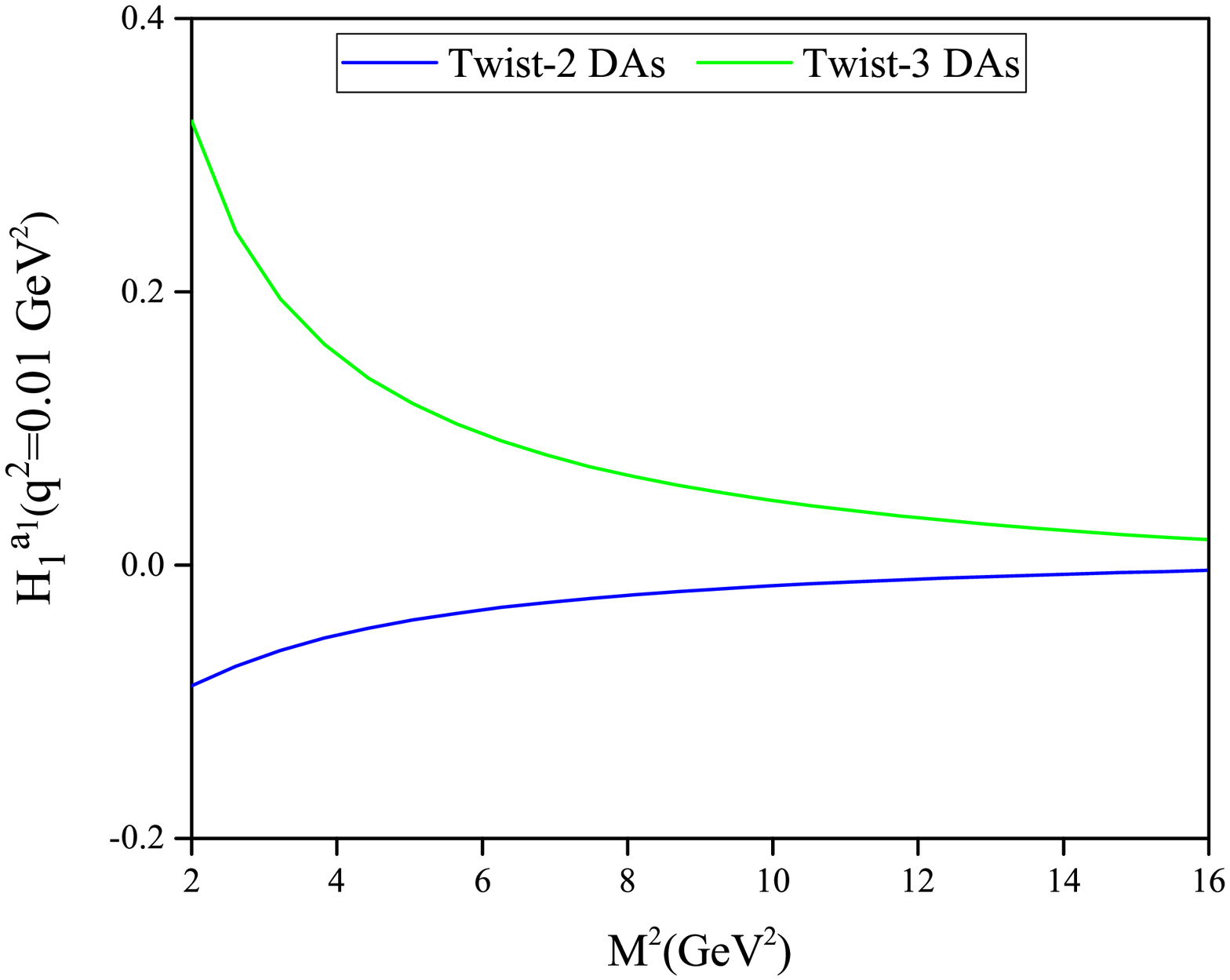}
\includegraphics[width=5.5cm,height=5.5cm]{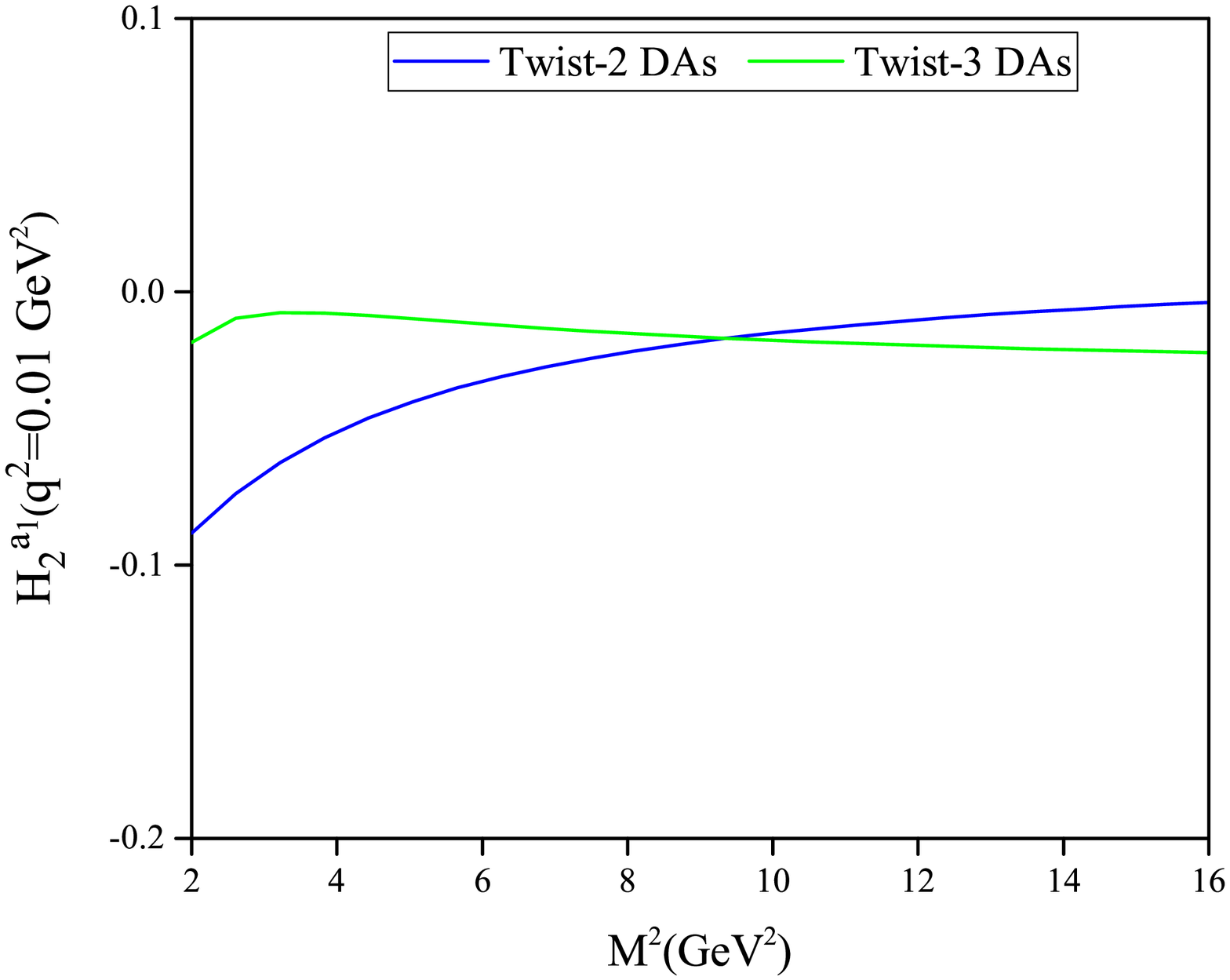}
\caption{ The contributions of
twist-2 and twist-3 distribution amplitudes in  the $D^0 \to a_1^{-}$ helicity form factors on $M^{2}$ and $s_{0}=7~ \rm{GeV}^{2}$. The values of $q^2$ are taken as Fig. \ref{Fb}.
}\label{Ft}
\end{figure}

\begin{figure}[th]
\includegraphics[width=5.5cm,height=5.5cm]{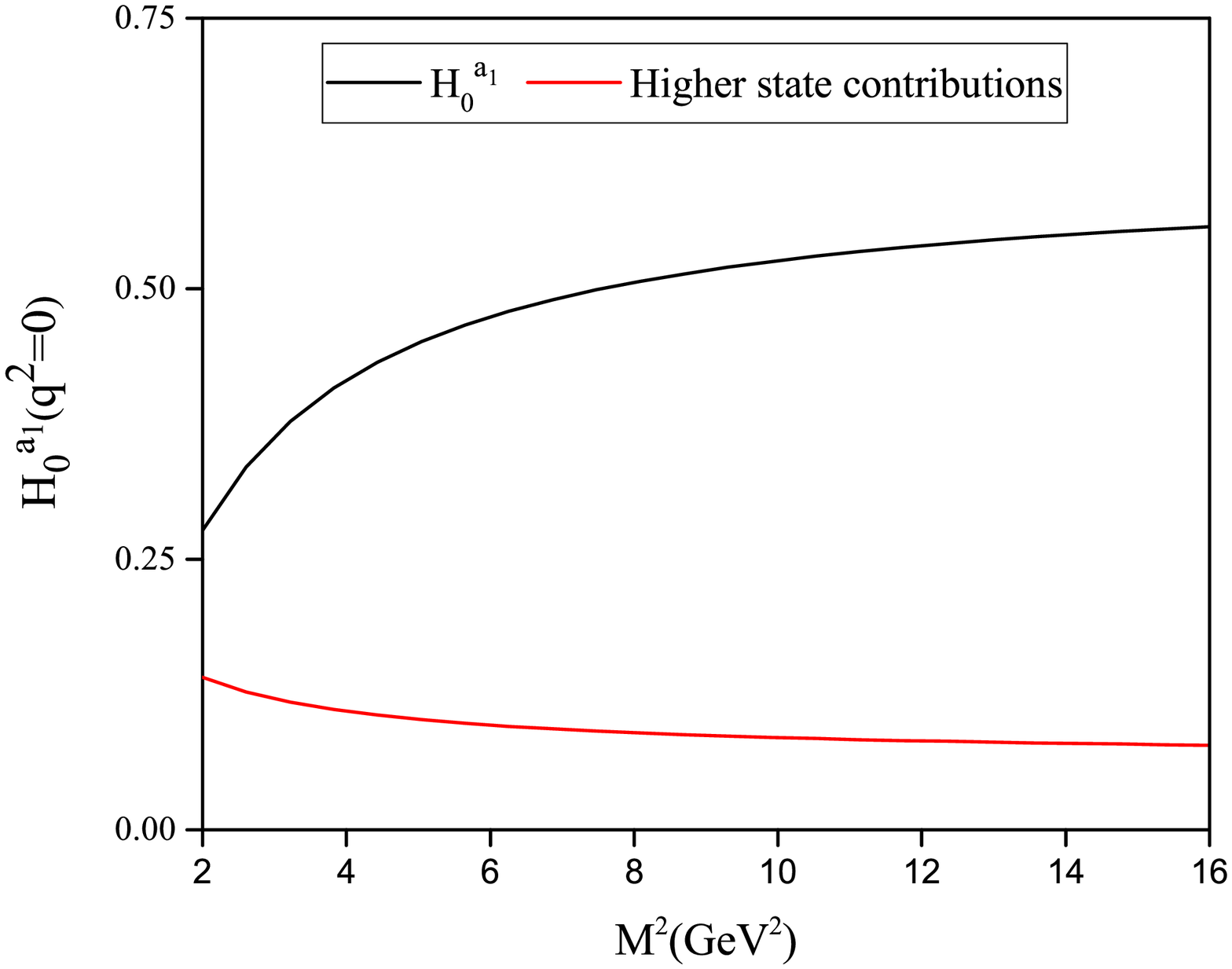}
\includegraphics[width=5.5cm,height=5.5cm]{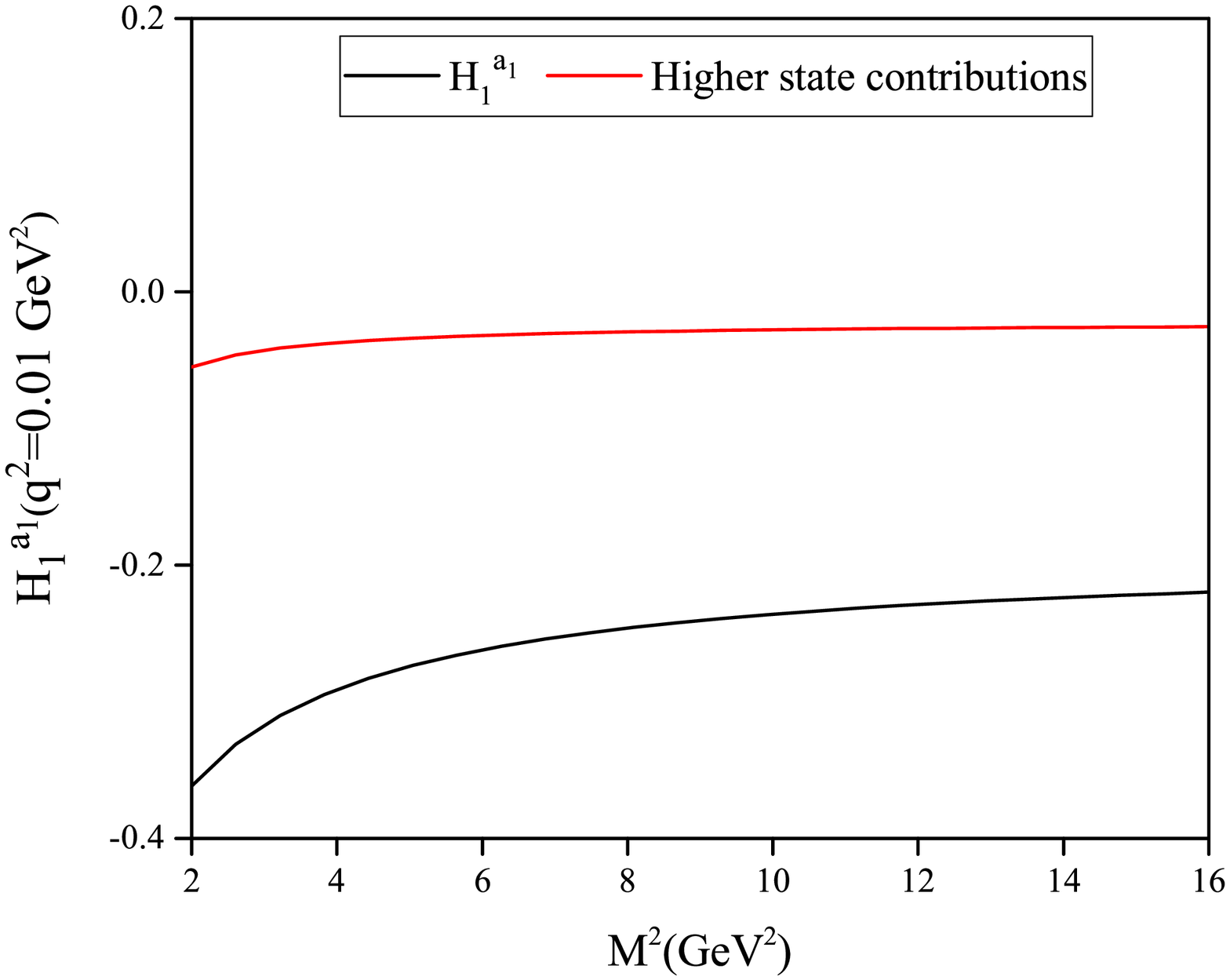}
\includegraphics[width=5.5cm,height=5.5cm]{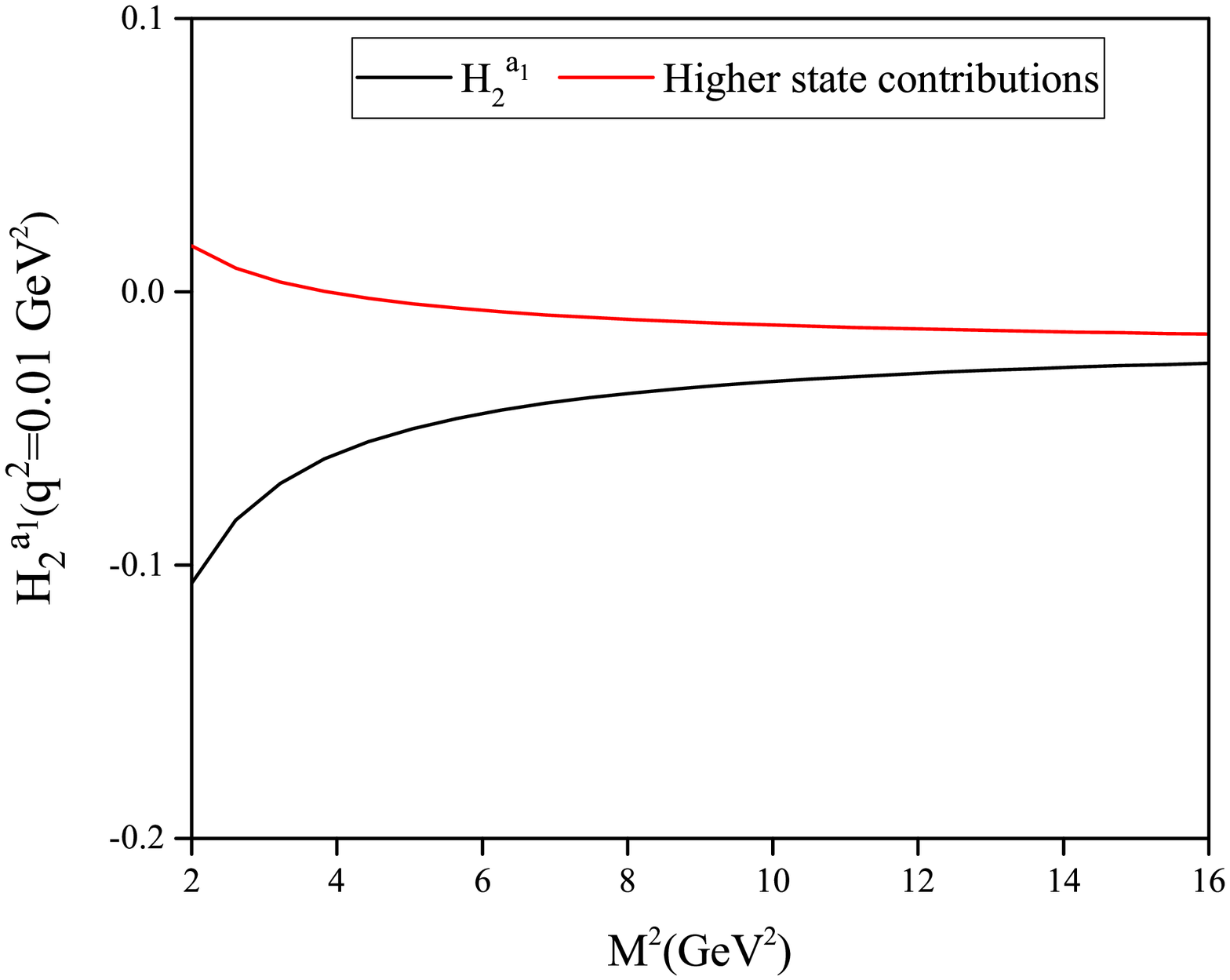}
\caption{$D^0 \to a_1^{-}$ helicity form factors as a function of  $M^{2}$ for  $s_{0}=7~ \rm{GeV}^{2}$ as well as the higher states contributions  in these form factors.  The values of $q^2$ are chosen as Fig. \ref{Fb}.}\label{Fh}
\end{figure}

The contributions of twist-2 and twist-3 distribution amplitudes and  higher states
 in the $D^0 \to a_1^{-}$ helicity form factors, with respect to
$M^{2}$,  are displaced in  Figs. \ref{Ft} and \ref{Fh}.
 It can be observed that at the above-mentioned  interval from Borel mass,
the higher twist contributions as well as higher states, are suppressed. Our numerical analysis shows, that the contribution of
the  higher states is smaller than about $8\% $ of the total value.


Using all the input values and parameters,  the helicity form factors
can be evaluated  as a function of $q^2$. The values of $\mathcal{H}_{0}$ for aforementioned decays
 at the zero transferred momentum
square $q^2=0$ are presented in Table \ref{T1}. In this table, the contributions of twist-2  distribution amplitudes
are also reported. The main
uncertainty in ${\mathcal{H}}_{0}(q^2=0)$ comes from  $c$ quark mass
$m_{c}$ and $\Phi_{\perp}$ light cone distribution amplitude.

\begin{table}
\begin{ruledtabular}
\begin{tabular}{ccccccc}
process&${\mathcal{H}}_{0}(q^2=0)$ & Twist-2  &process&${\mathcal{H}}_{0}(q^2=0)$ & Twist-2 \\
\hline
$D^0\to a^{-}_{1}\ell^+ \nu$& $0.67 ^{+0.26}_{-0.08} $&$0.56 ^{+0.21}_{-0.05}$  &
$D^0\to b^{-}_{1}\ell^+ \nu$&$-0.76 ^{+0.22}_{-0.19} $& $-0.62 ^{+0.16}_{-0.10}$   \\
$D^+\to a^{0}_{1}\ell^+ \nu$&$0.46 ^{+0.18}_{-0.05}$&  $0.38 ^{+0.14}_{-0.03}$    &
$D^+\to b^{0}_{1}\ell^+ \nu$&$-0.53 ^{+0.15}_{-0.13} $& $-0.43 ^{+0.11}_{-0.10} $  \\
$D\to K_{1A}\ell^+ \nu$&$0.51 ^{+0.20}_{-0.04}$& $0.40 ^{+0.15}_{-0.03}$   &
$D\to K_{1B}\ell^+ \nu$&$-0.95 ^{+0.25}_{-0.21}$&  $-0.83 ^{+0.21}_{-0.15}$ \\
$D_{s}\to K_{1A}\ell^+ \nu$&$0.31 ^{+0.14}_{-0.12}$&  $0.21 ^{+0.11}_{-0.08}$ &
$D_{s}\to K_{1B}\ell^+ \nu$&$-0.40 ^{+0.15}_{-0.18}$& $-0.32 ^{+0.11}_{-0.12}$  \\
\end{tabular}
\caption{Helicity form factor ${\mathcal{H}}_{0}$ as well as contribution of twist-2 distribution
amplitudes of the $D^{+} \to a^{0}_{1} (b^{0}_{1}) ~\ell^+ \nu$, $D^{0} \to
a^{-}_{1} (b^{-}_{1} )~ \ell^+ \nu$, $D\to K_{1A} (K_{1B})~\ell^+ \nu$ and $D_{s}\to K_{1A} (K_{1B})~
\ell^+ \nu$ decays at $q^2=0$.} \label{T1}
\end{ruledtabular}
\end{table}

In order to extend LCSR prediction to the whole
physical region, $ m_\ell^2
\le q^2 \le (m_{D_{(s)}}-m_{A})^2$,
we use the series
expansion given in \cite{Bharucha2010} as:
\begin{eqnarray}
\mathcal{H}_{0}^{A}(q^2) &=&\frac{1}{z(q^2, m_{D_{(s)}^{r}}^2)\sqrt{z(q^2,t_{-})}\,\phi(q^2) } \sum_{k=0,1} a_k^{A, 0} z^k(q^2, t_{0}) , \label{ff1} \\
\mathcal{H}_{1}^{A}(q^2) &=&\frac{\sqrt{-z(q^2, 0)}}{z(q^2, m_{D_{(s)}^{r}}^2)\,\phi(q^2)} \sum_{k=0,1} a_k^{A, 1} z^k(q^2, t_{0}) , \label{ff2}  \\
\mathcal{H}_{2}^{A}(q^2) &=&\frac{\sqrt{-z(q^2, 0)}}{z(q^2, m_{D_{(s)}^{r}}^2)\sqrt{z(q^2,t_{-})}\,\phi(q^2)  } \sum_{k=0,1} a_k^{A, 2} z^k(q^2, t_{0}) ,\label{ff3}
\end{eqnarray}
where
\begin{eqnarray}
&z(q^2, t)=\frac{\sqrt{t_+ - q^2}-\sqrt{t_+ - t}}{\sqrt{t_+ - q^2}+\sqrt{t_+ - t}},\\
&\sqrt{-z(q^2, 0)}=\sqrt{q^2}/m_{D_{(s)}^{r}},
\end{eqnarray}

where $t=t_{0},~ t_{-},~ m_{D_{(s)}}$ with $t_\pm=(m_{D_{(s)}}\pm m_{A})^2$ and $t_0=t_+(1-\sqrt{1-t_-/t_+})$.
Moreover, $D_{(s)}^{r}$ shows the resonance states are given in Table \ref{Tm}. The
function $\phi(q^2)$ is given by \cite{Arnesen2005}:
\begin{eqnarray}
\phi(q^2)=\sqrt{\frac{3t_{+}t_{-}}{32\pi\chi_{0}}}\,\left(\frac{z(q^2, 0)}{-q^2}\right)^2\,\left(\frac{z(q^2, t_{0})}{t_{0}-q^2}\right)^{-0.5}\,\left(\frac{z(q^2, t_{-})}{t_{-}-q^2}\right)^{-0.25}\,\left(\frac{(t_{+}-q^2)^2}{t_{+}-t_{0}}\right)^{0.25},
\end{eqnarray}
where $\chi_{0}$ has been calculated  using OPE and is given by \cite{Bharucha2010}:
\begin{eqnarray}\label{chi}
\chi_{0}=\frac{1+0.751\alpha_{s}(m_{c})}{8\pi^2}
\end{eqnarray}
It should be noted that for the functions  $\sqrt{z(q^2,t_{-})}$ and $\phi(q^2)$  the replacement $m_{D_{(s)}}\to m_{D_{(s)}^{r}}$ must be made.
For the series expansion parameterizations \ref{ff1}, \ref{ff2} and \ref{ff3}, the unitarity constraints are obtained as
\cite{Bharucha2010}:
\begin{eqnarray}\label{cons}
\sum_{k=0,1}\left\{(a_k^{A, 0})^2+(a_k^{A, 1})^2+(a_k^{A, 2})^2 \right\}\leq 1
\end{eqnarray}
We use   parameter $\Delta$  defined as:
\begin{eqnarray}
\Delta=\frac{\sum_{q^2}\left|\mathcal{H}^{A}_{\sigma}(q^2)-\mathcal{H}^{A, \rm fit}_{\sigma}(q^2)\right|} {\sum_{q^2}\left|\mathcal{H}^{A}_{\sigma}(q^2)\right|}\times 100, \label{delta}
\end{eqnarray}
where $0\leq~q^2~\leq
(m_{D_{(s)}}-m_{A})^2/2$ to estimate quality of fit for each helicity form factor.
Table \ref{T2} includes the
values of $a_{1}^{\sigma}$, $a_{2}^{\sigma}$ and $\Delta$ for the helicity form factors of the semileptonic decays.
For  these results  all the input
parameters are set to be their central values.
As it can be seen from the values of $\Delta$  parameters, are reported in \ref{T2}, the fit functions \ref{ff1}, \ref{ff2} and \ref{ff3} cover the LCSR predictions for the helicity form factors.
\begin{table}[th]
\caption{Values of $b_{0}$, $b_{1}$ and $b_{2}$ related to
$F^{(1)}(q^2)$ for the fitted form factors of $D_{(s)}\to a_{1},
b_{1}, K_{1A}$ and $K_{1B}$ transitions.} \label{T2}
\begin{ruledtabular}
\begin{tabular}{cccc|cccc}
$\mbox{Form factor}$& ${a_{1}}$&$a_{2}$&$\Delta$ &$\mbox{Form factor}$& ${a_{1}}$&$a_{2}$&$\Delta$\\
\hline
$\mathcal{H}_{0}^{D^0\to a^{-}_{1}}$& ${0.05}$&$-0.95$&$0.36$ &$\mathcal{H}_{0}^{D^0\to b^{-}_{1}}$& ${-0.10}$&$0.49$&$0.32$\\
$\mathcal{H}_{1}^{D^0\to a^{-}_{1}}$& ${-0.07}$&$-0.56$&$0.26$ &$\mathcal{H}_{1}^{D^0\to b^{-}_{1}}$&${0.12}$&$-0.54$&$0.17$\\
$\mathcal{H}_{2}^{D^0\to a^{-}_{1}}$& ${-0.12}$&$0.83$&$0.24$ &$\mathcal{H}_{2}^{D^0\to b^{-}_{1}}$&${0.10}$&$-0.87$&$0.50$\\
$\mathcal{H}_{0}^{D^+\to a^{0}_{1}}$& ${0.03}$&$-0.67$&$0.35$ &$\mathcal{H}_{0}^{D^+\to b^{0}_{1}}$& ${-0.07}$&$0.34$&$0.31$\\
$\mathcal{H}_{1}^{D^+\to a^{0}_{1}}$& ${-0.04}$&$-0.39$&$0.26$ &$\mathcal{H}_{1}^{D^+\to b^{0}_{1}}$&${0.08}$&$-0.38$&$0.16$\\
$\mathcal{H}_{2}^{D^+\to a^{0}_{1}}$& ${-0.09}$&$0.59$&$0.24$ &$\mathcal{H}_{2}^{D^+\to b^{0}_{1}}$& ${0.07}$&$-0.61$&$0.48$\\
$\mathcal{H}_{0}^{D\to K_{1A}}$& ${0.12}$&$-0.54$&$0.57$ &$\mathcal{H}_{0}^{D\to K_{1B}}$& ${-0.04}$&$0.68$&$0.83$\\
$\mathcal{H}_{1}^{D\to K_{1A}}$& ${-0.09}$&$-0.85$&$0.30$ &$\mathcal{H}_{1}^{D\to K_{1B}}$& ${0.16}$&$-0.71$&$0.57$\\
$\mathcal{H}_{2}^{D\to K_{1A}}$& ${-0.02}$&$0.08$&$0.29$ &$\mathcal{H}_{2}^{D\to K_{1B}}$& ${0.16}$&$-0.90$&$0.54$\\
$\mathcal{H}_{0}^{D_s\to K_{1A}}$& ${0.02}$&$0.85$&$0.17$ &$\mathcal{H}_{0}^{D_s\to K_{1B}}$& ${-0.05}$&$0.72$&$0.22$\\
$\mathcal{H}_{1}^{D_s\to K_{1A}}$& ${-0.07}$&$-0.76$&$0.05$ &$\mathcal{H}_{1}^{D_s\to K_{1B}}$& ${0.12}$&$-0.82$&$0.19$\\
$\mathcal{H}_{2}^{D_s\to K_{1A}}$& ${-0.01}$&$-0.14$&$0.05$ &$\mathcal{H}_{2}^{D_s\to K_{1B}}$& ${0.04}$&$-0.77$&$0.11$
\end{tabular}
\end{ruledtabular}
\end{table}

\begin{figure}[th]
\includegraphics[width=5.5cm,height=5.5cm]{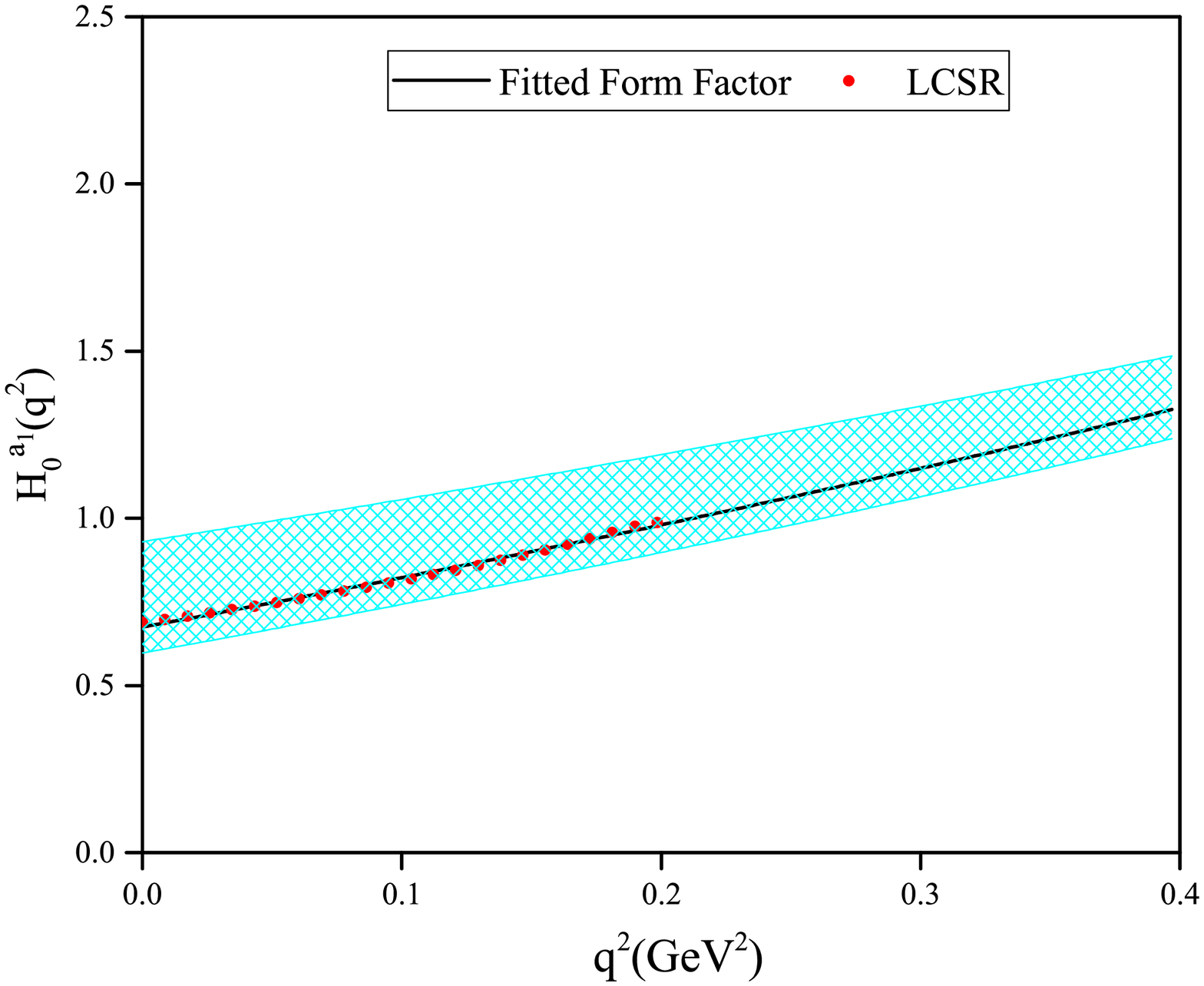}
\includegraphics[width=5.5cm,height=5.5cm]{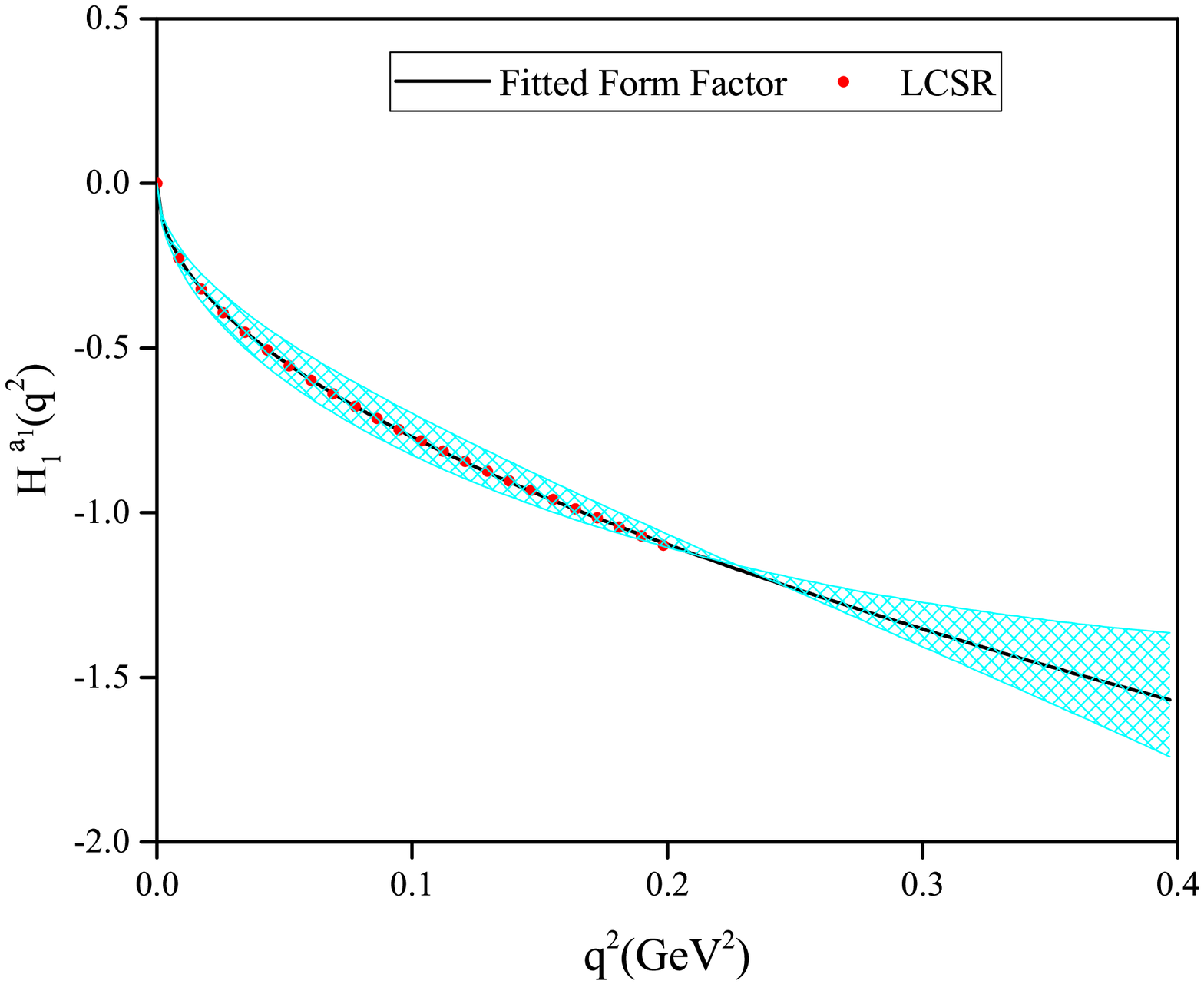}
\includegraphics[width=5.5cm,height=5.5cm]{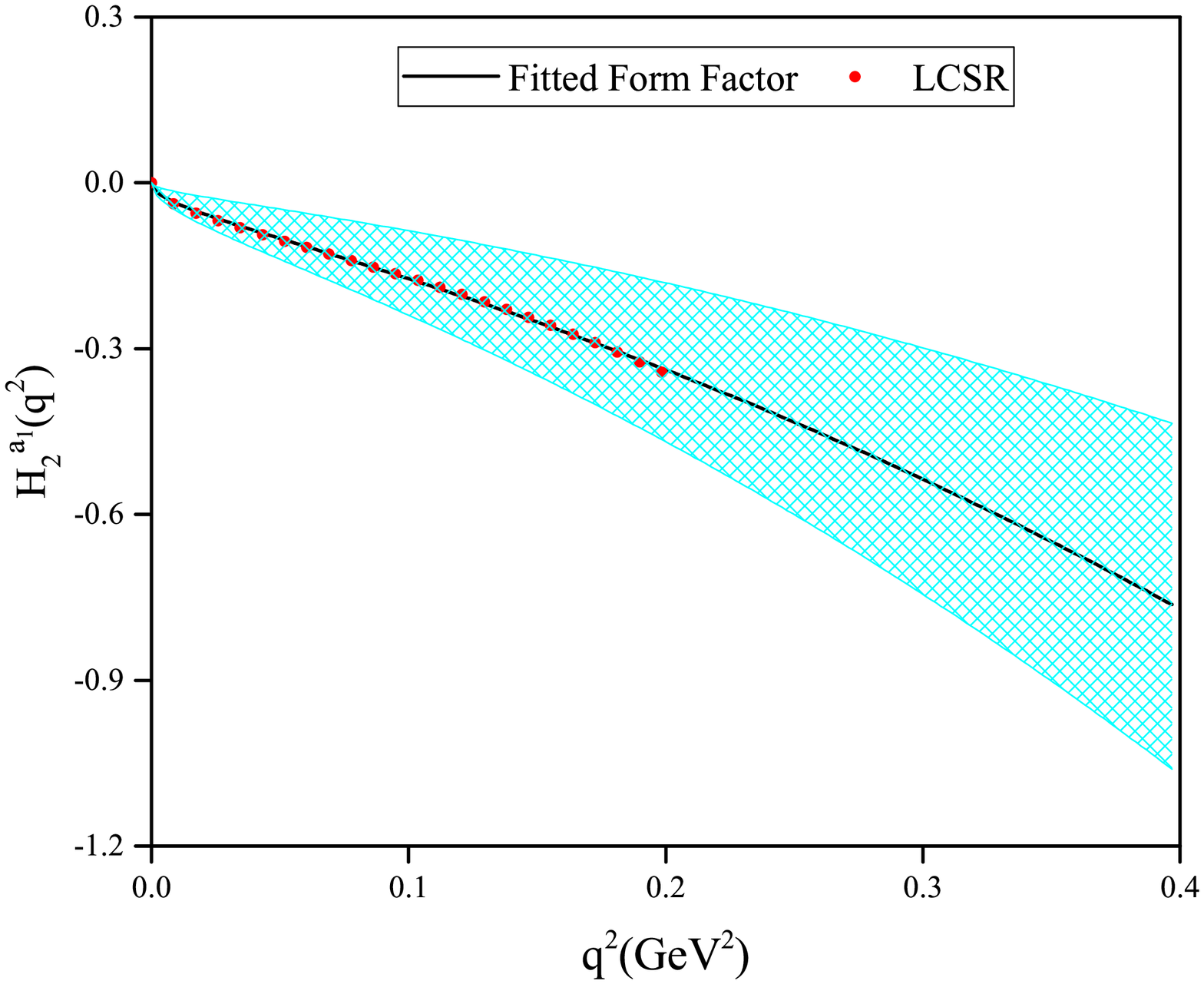}
\includegraphics[width=5.5cm,height=5.5cm]{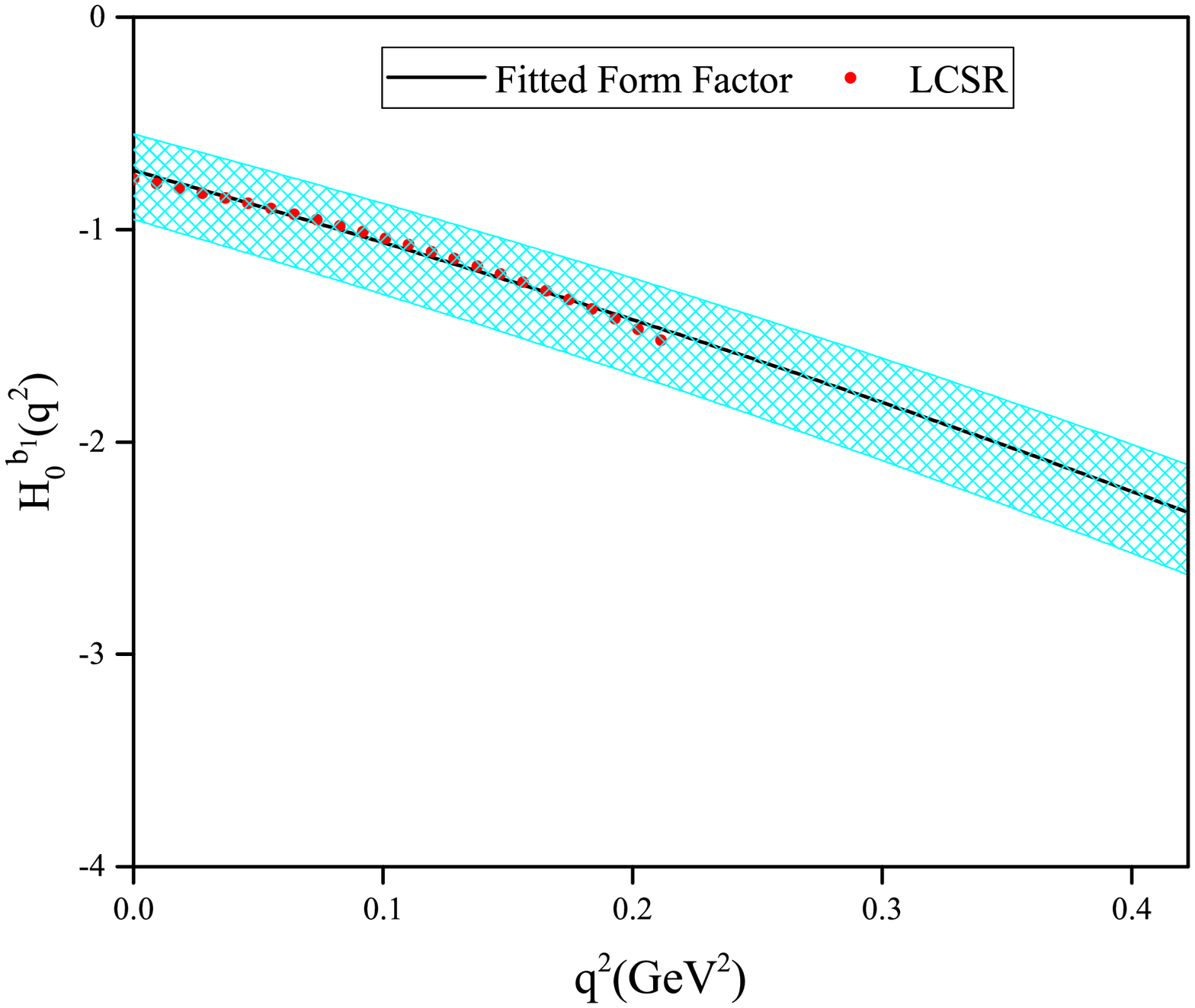}
\includegraphics[width=5.5cm,height=5.5cm]{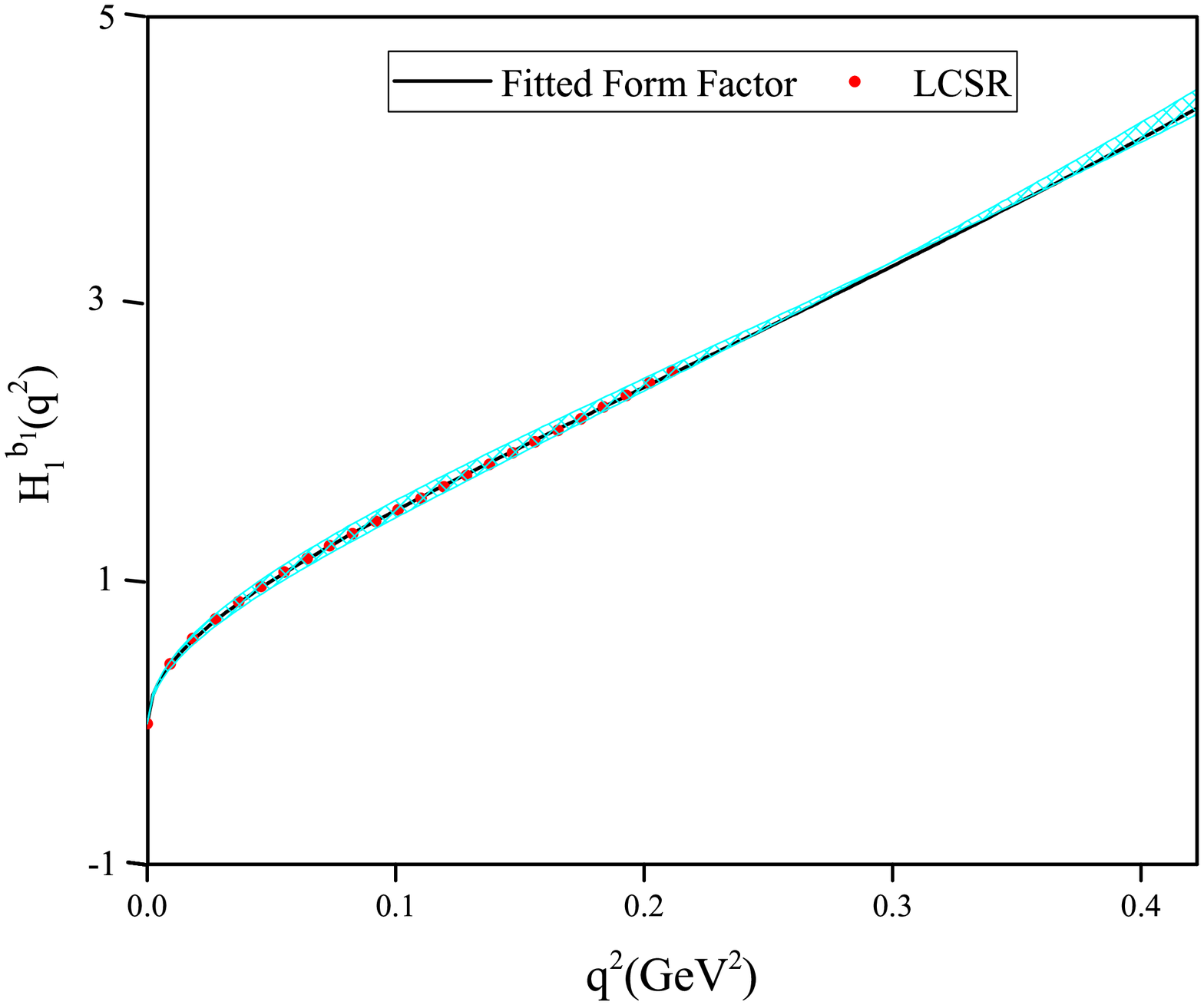}
\includegraphics[width=5.5cm,height=5.5cm]{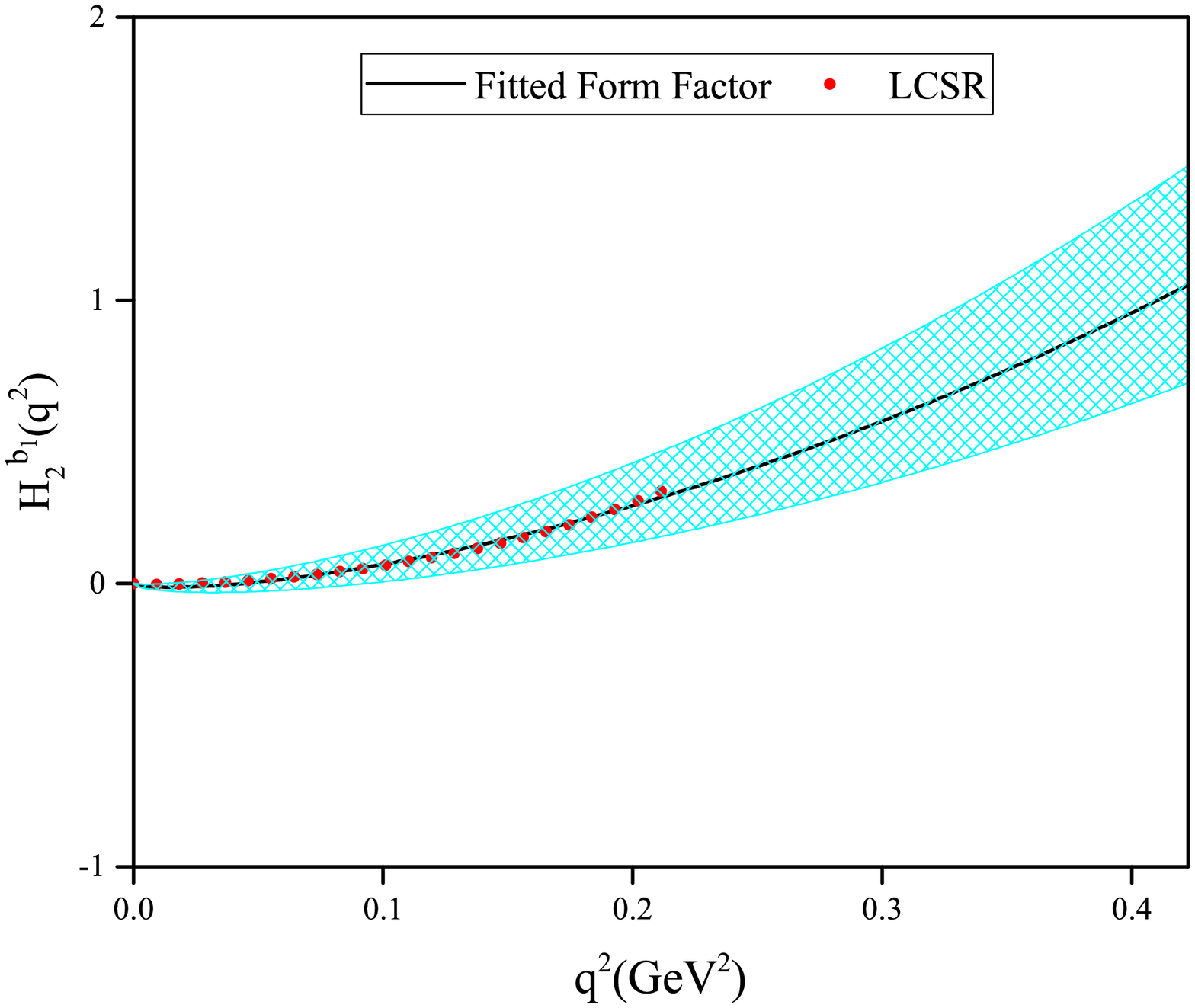}
\caption{$D^0 \to a_1^{-}$ and $D^0 \to b_1^{-}$ helicity form factors as a function of  $q^{2}$. Circles show the results of the LCSR while, black lines show the fitted form factors in the  whole
physical regions. The shaded bands stand for the results correspond the upper and lower values of the input parameters.}\label{ff}
\end{figure}

The dependence of the form factors $\mathcal{H}_{0}$, $\mathcal{H}_{1}$ and $\mathcal{H}_{2}$ for
$D^0\to a^{-}_{1}$  and $D^0\to b^{-}_{1}$ transitions on $q^2$ are plotted in Fig. \ref{ff}.
In these plots, the LCSR results and the  fitted form factors are displaced with circles
and black lines, respectively. Moreover, the shaded regions are obtained using upper and lower values of the input parameters.

\subsection{Analysis of the branching ratios }
\begin{figure}[th]
\includegraphics[width=7cm,height=7cm]{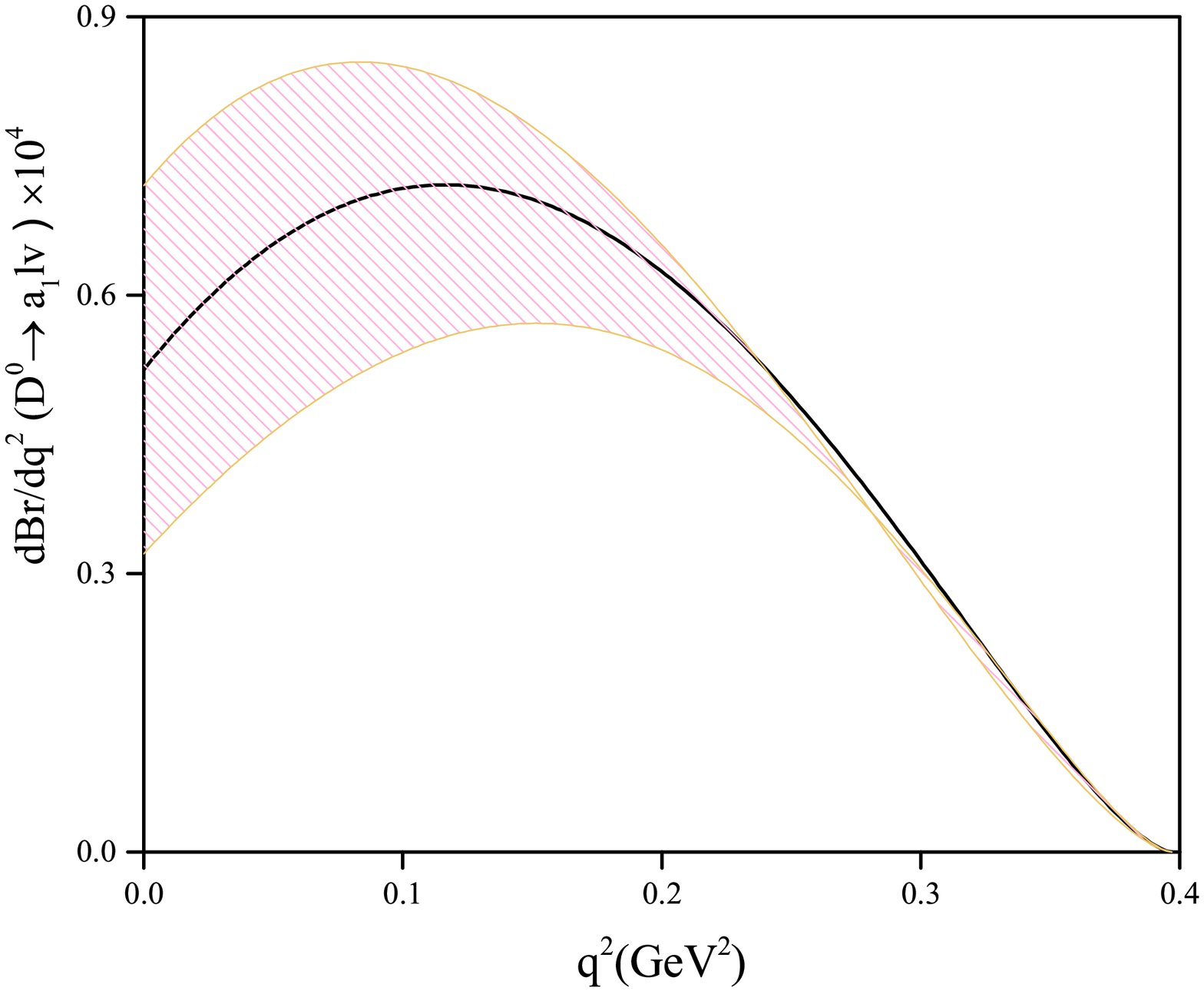}
\includegraphics[width=7cm,height=7cm]{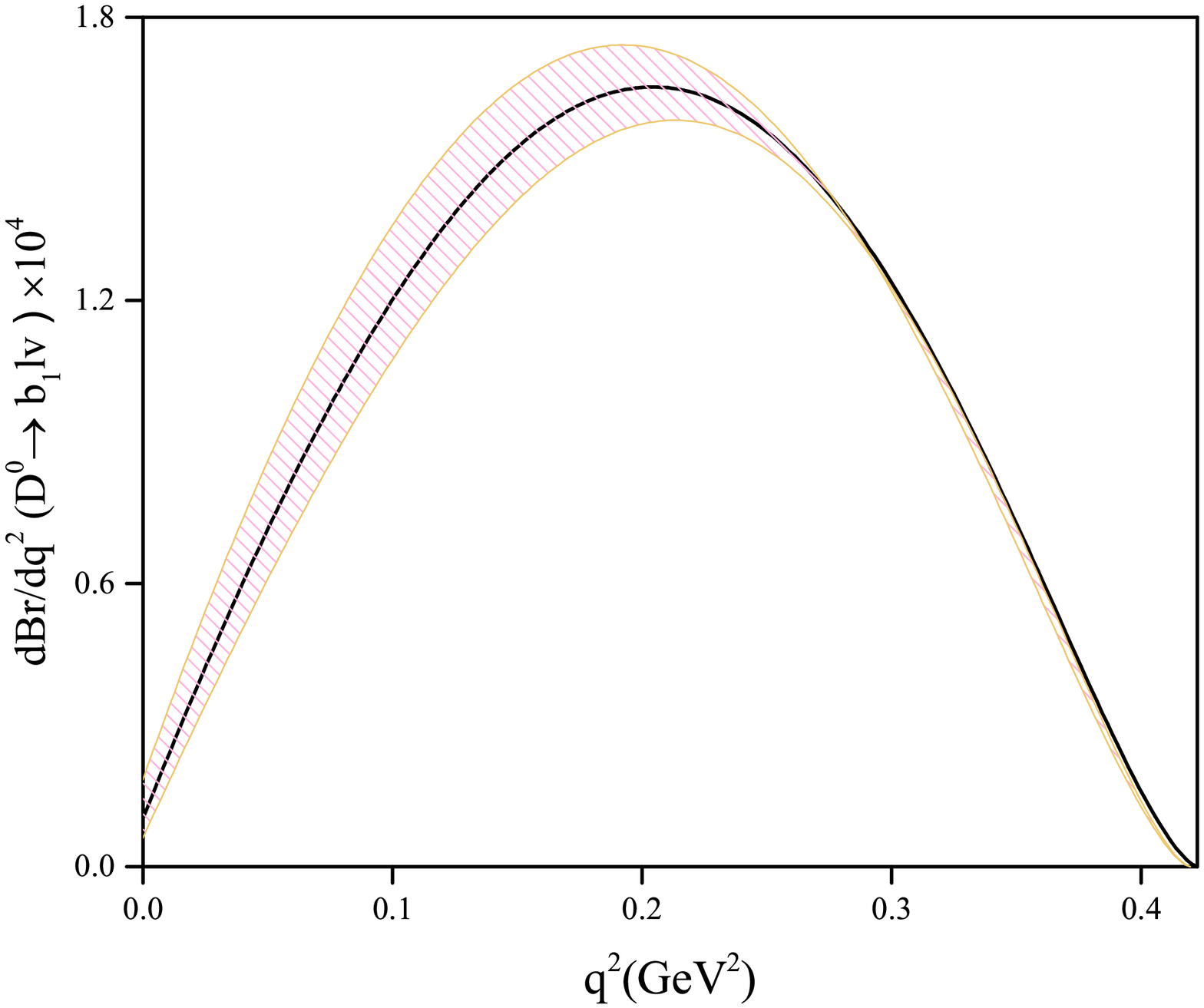}
\caption{The differential branching ratio of $D^0 \to a_1^{-}$ and $D^0 \to b_1^{-}$
decays as a function of $q^2$. The shaded intervals show  the results obtained using the upper and lower values of the input parameters.}\label{bra1b1}
\end{figure}

\begin{figure}[th]
\includegraphics[width=10cm,height=6.5cm]{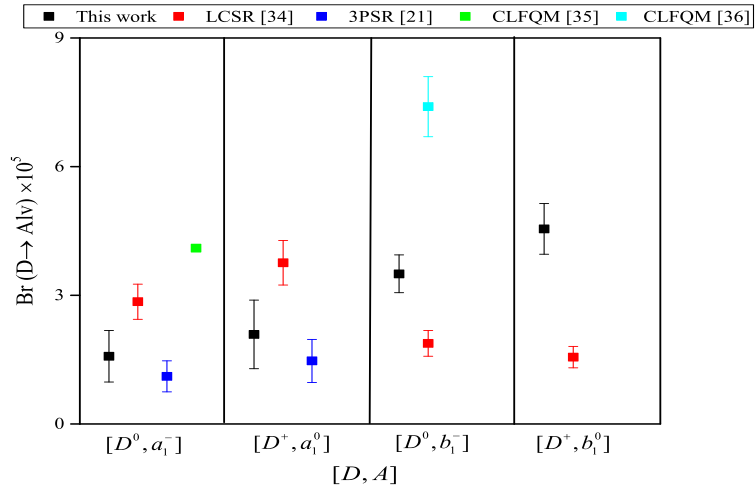}
\caption{Our predictions for branching ratio values of the semileptonic $D^{0} \to a_{1}^{-}(b_{1}^{-})
\ell \nu $ and $D^+\to a^{0}_{1}( b^{0}_{1})\ell \nu$ decays. The results of the other methods, estimated using transition form factors, such as LCSR, 3PSR and CLFQM
are also reported.   }\label{br1}
\end{figure}

Now, we are ready to estimate the branching
ratio values for the semileptonic $D_{(s)} \to A \ell \nu$ decays.
The  differential decay width of  considered semileptonic decays is
evaluated in SM as:
\begin{eqnarray}
\frac{d\Gamma (D_{(s)}\rightarrow A \ell \nu)}{dq^2}=\frac{ {\sqrt{\lambda}}\,
G_F^2\, \left| V_{cq'}\right| ^{2}} {192\,\pi^3\,
m_{D_{(s)}}^3}\Bigg\{[\mathcal{H}^{A}_{0}(q^2)]^2 +[\mathcal{H}^{A}_{1}(q^2)]^2+[\mathcal{H}^{A}_{2}(q^2)]^2\Bigg\}, \label{difftot}
\end{eqnarray}
where   $V_{cq'}=V_{cd}(V_{cs})$ is used
for $c\to d(s)\, \ell \nu$ transition. To calculate the branching ratios, the total mean life time
$\tau_{D^0}=0.41 $, $\tau_{D^+}=1.04$ and
$\tau_{D^{+}_s}=0.50 $ ps \cite{pdg} are used for the $D_{(s)}$ states.
The differential branching ratios of  $ D^0 \to a^{-}_{1} (b^{-}_{1})
\ell \nu $  with their uncertainly regions, are plotted with respect to $q^2$ in Fig. \ref{bra1b1}.
Moreover, our  results for the branching ratio values of the semileptonic decays $D^{0} \to a_{1}^{-}(b_{1}^{-})
\ell \nu $ and $D^+\to a^{0}_{1}( b^{0}_{1})\ell \nu$ decays   as well as  the estimations of  the other
approaches are presented in Fig. \ref{br1}.  The predictions of LCSR, 3PSR and CLFQM are calculated by using transition form factors.

\begin{figure}[th]
\includegraphics[width=7cm,height=6.5cm]{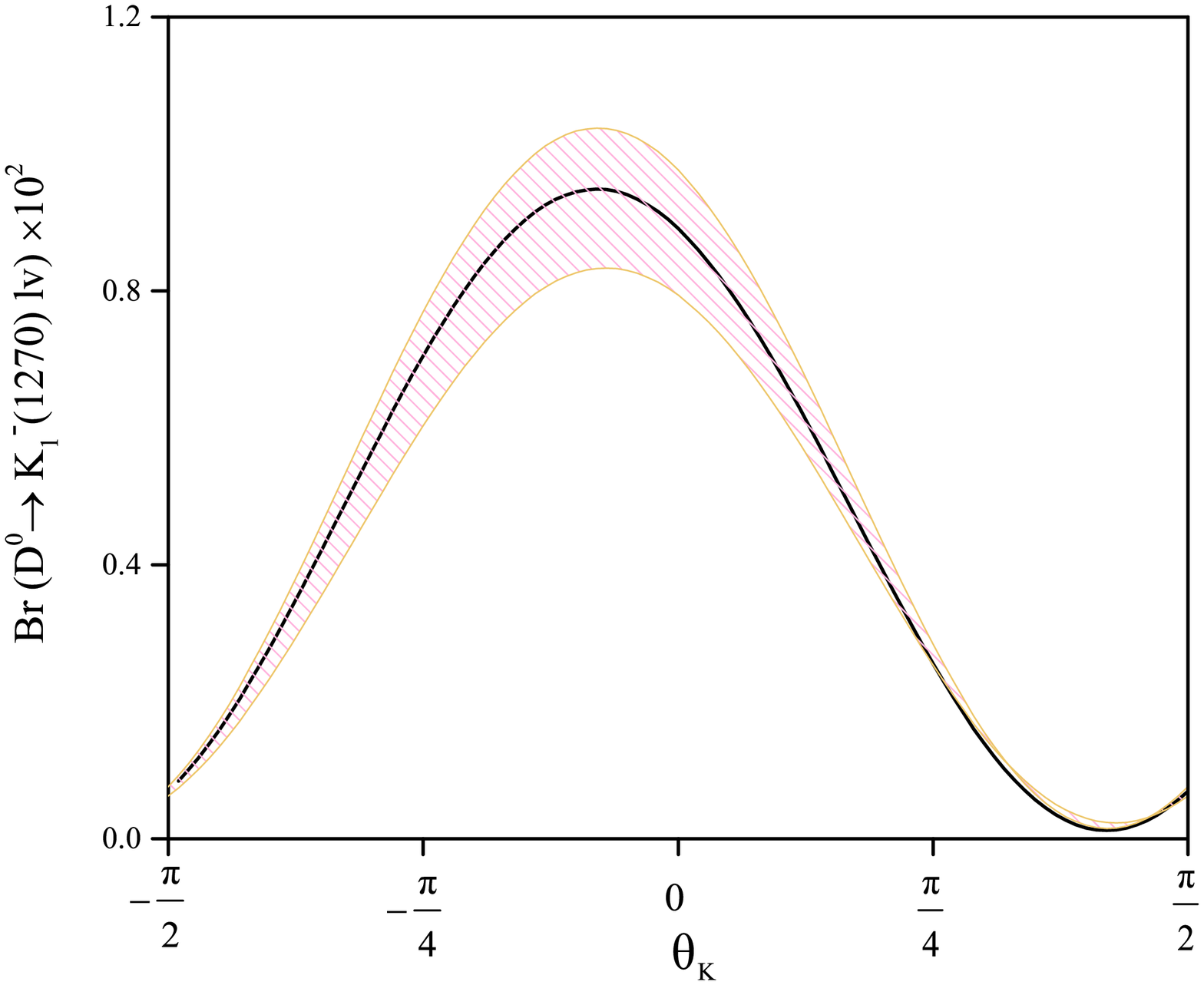}
\includegraphics[width=7cm,height=6.5cm]{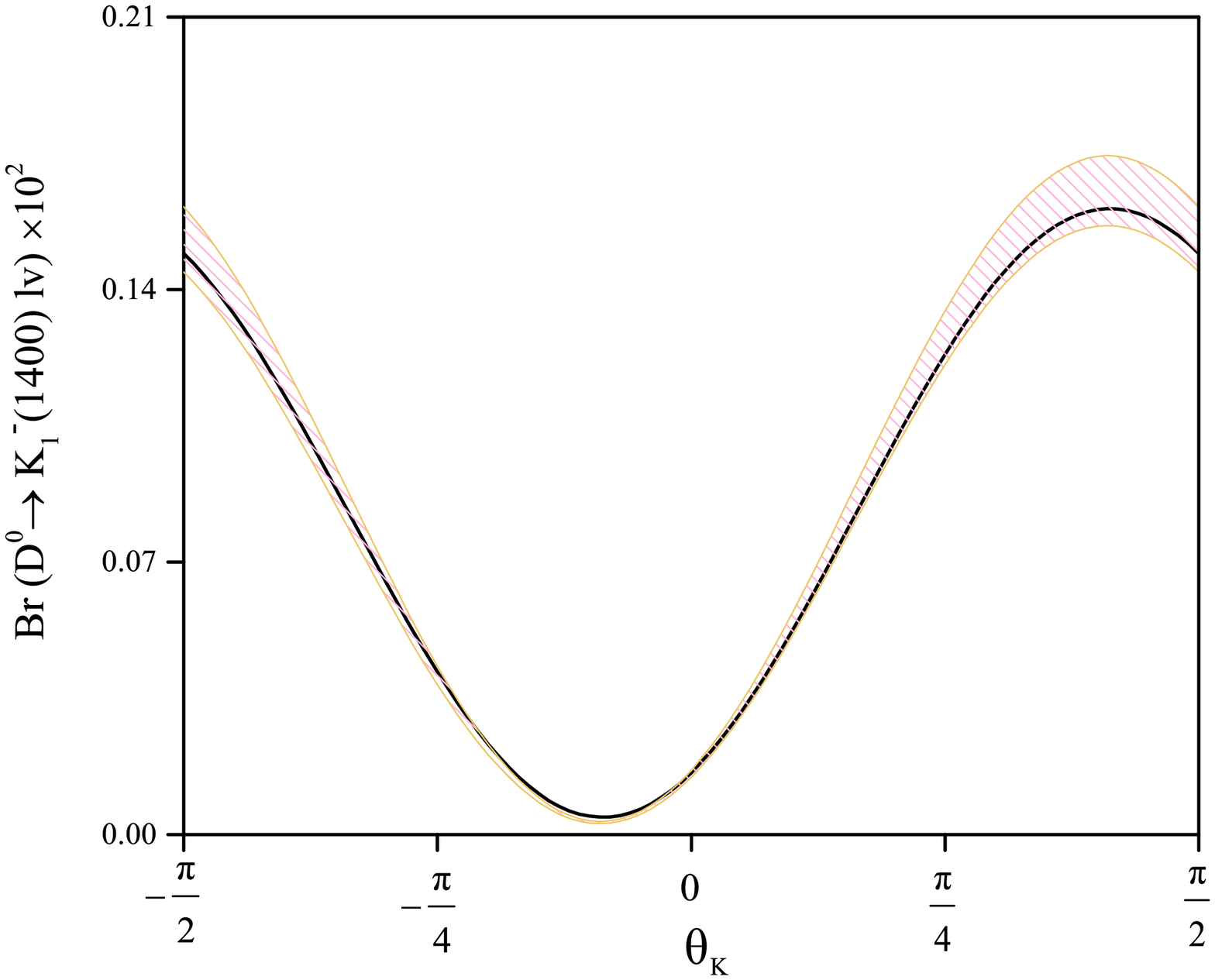}
\includegraphics[width=7cm,height=6.5cm]{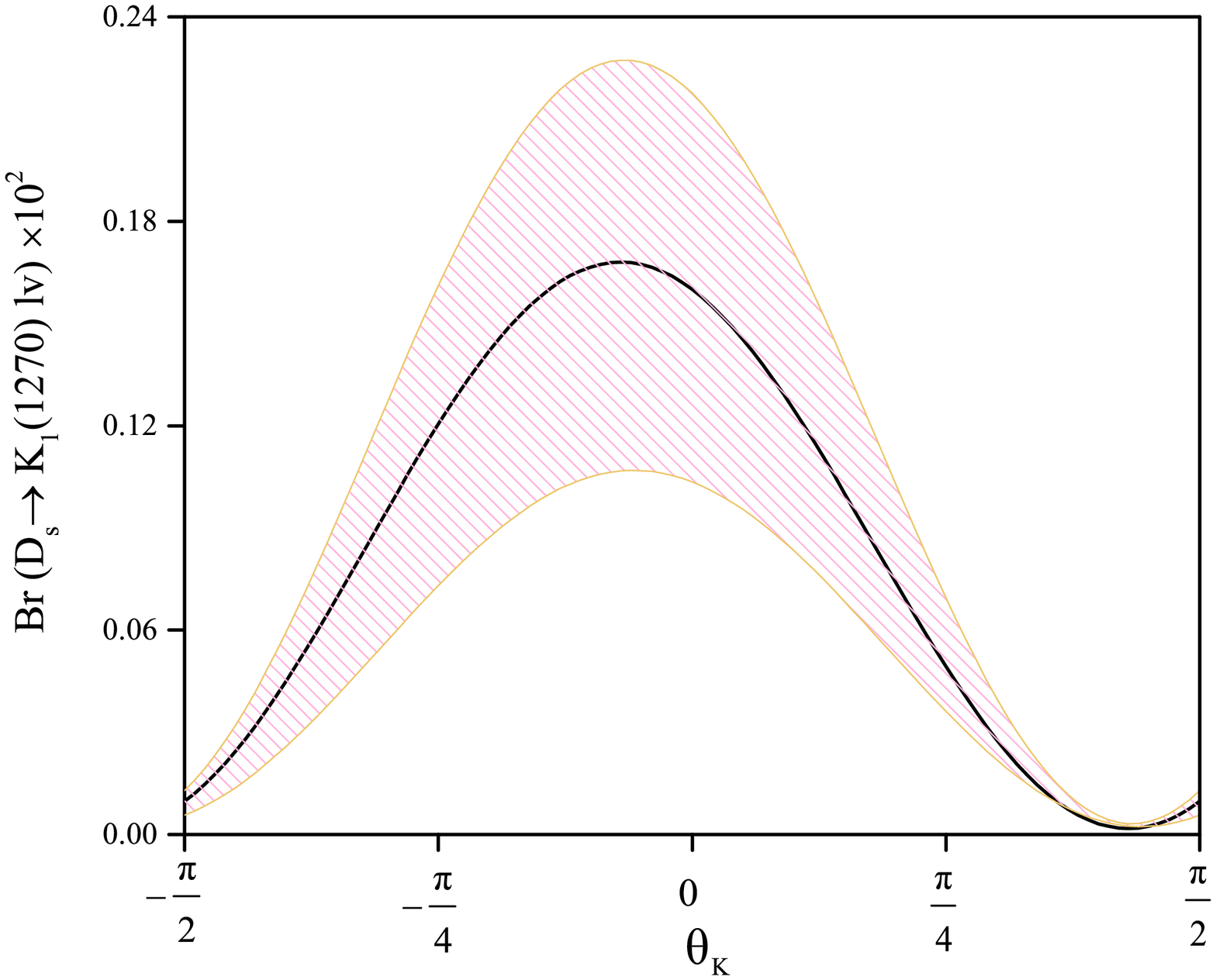}
\includegraphics[width=7cm,height=6.5cm]{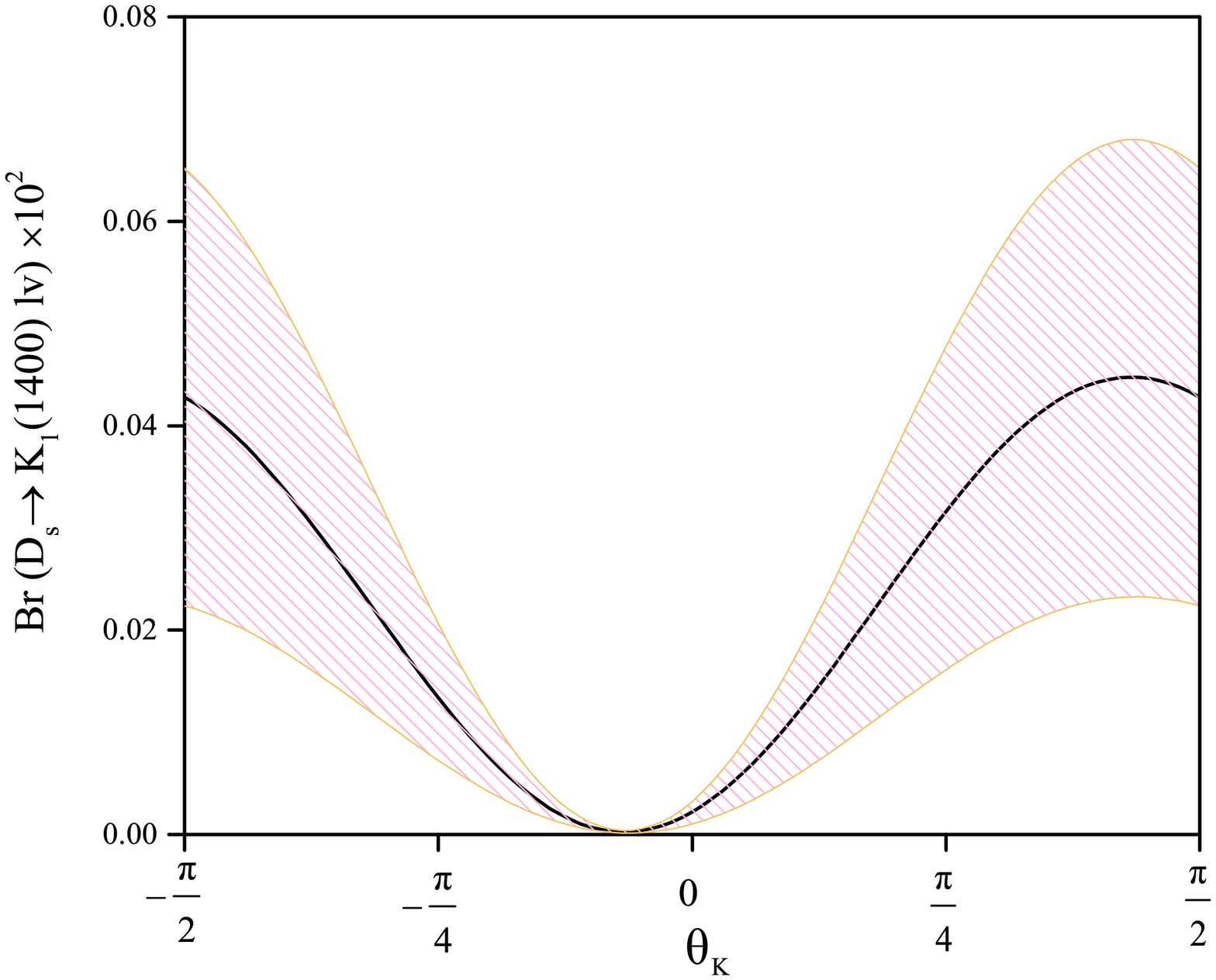}
\caption{ The $\theta_{K}$ dependence of differential branching ratios of  the semileptonic $D_{(s)} \to K_{1}(1270) \ell \nu$ and  $D_{(s)} \to K_{1}(1400) \ell \nu$   transitions with their uncertainly bands.}\label{brk1}
\end{figure}

\begin{figure}[th]
\includegraphics[width=8cm,height=7cm]{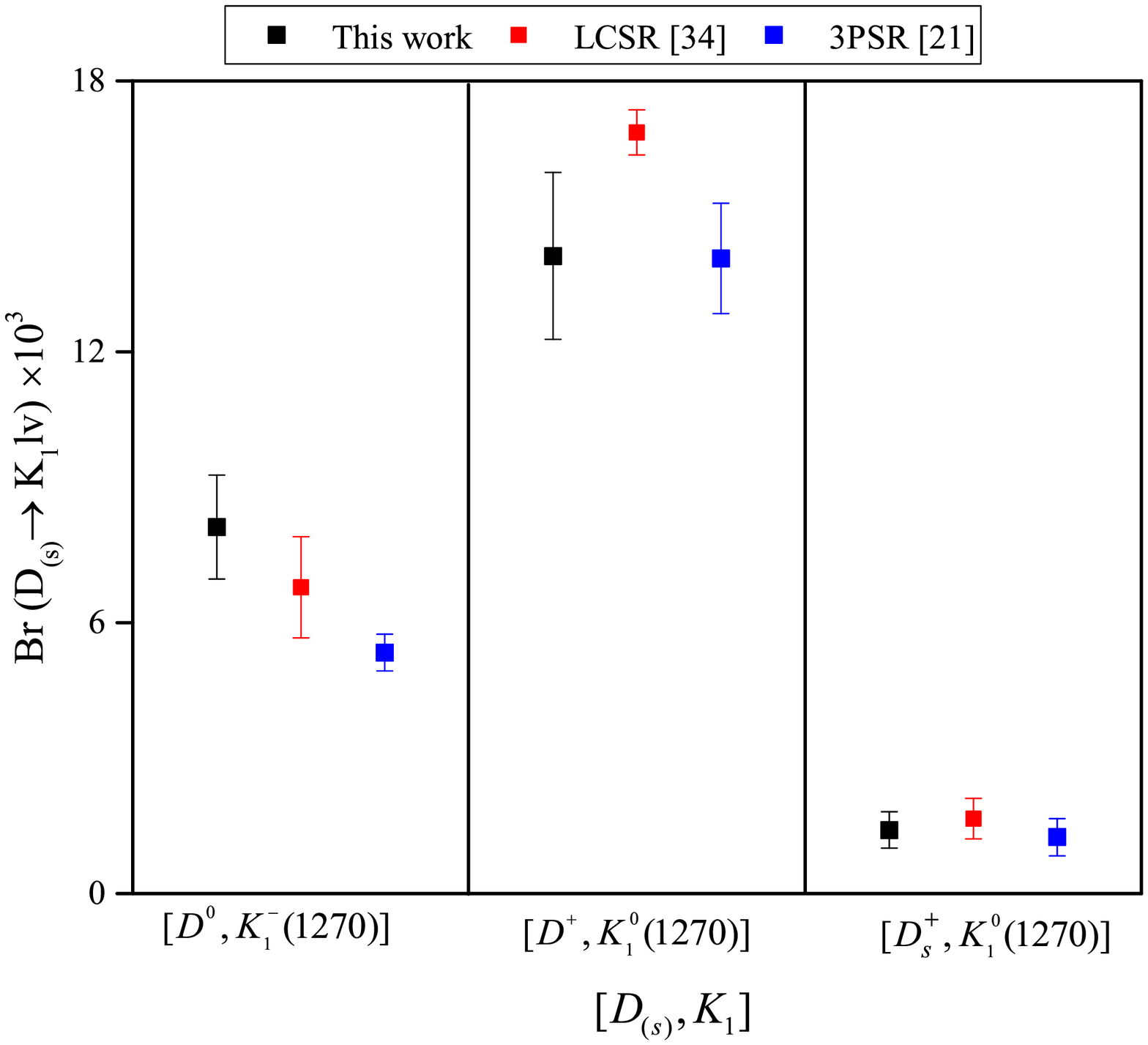}
\includegraphics[width=8cm,height=7cm]{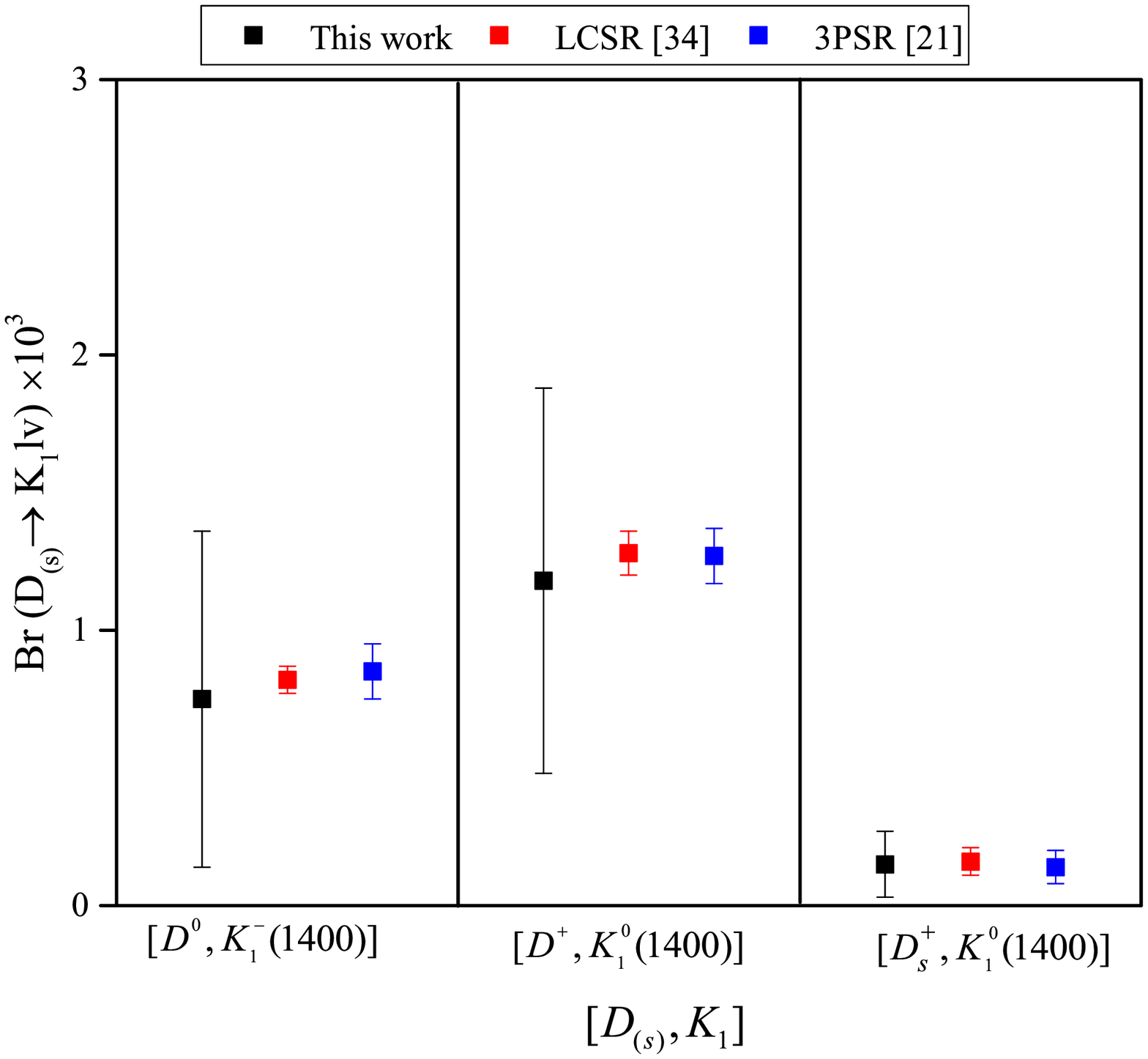}
\caption{Theatrical values for the branching ratio of the semileptonic $D_{(s)} \to K_{1}$ with $K_{1}=K_{1}(1270), K_{1}(1400)$ at $\theta_K ={-(34\pm13)}^{\circ}$. }\label{br2}
\end{figure}

The $\theta_{K}$ dependence of the branching ratio  values of $D_{(s)}$ decays into the physical states $ K_{1}(1270)$ and  $K_{1}(1400)$, are displaced in Fig. \ref{brk1}; and  comparison between our results and other theoretical technics at
 $\theta_K ={-(34\pm13)}^{\circ}$ are given in Fig. \ref{br2}.
The  $D^{+}\to K_{1}^{0}(1270)~e^{+}\nu_{e}$ decay is searched  at the BEPCII collider and
its decay branching fraction is determined to be $\mathcal{B}(D^{+}\to K_{1}^{0}(1270)~e^{+}\nu_{e})=(2.30\pm 0.69)$ \cite{Ablikim2019}.
Our branching ratio  of $D^{+}\to K_{1}^{0}(1270)~e^{+}\nu_{e}$ agrees with the experimental measurement
when $\theta_{K}=-(36.68\pm6.30)^{\circ}$.

In summary, we calculate the $D_{(s)}$ to  axial vector mesons $a_{1}^{-}$, $a_{1}^{0}$,
$b_{1}^{-}$, $b_{1}^{0}$, $K_{1}(1270)$ and $K_{1}(1400)$
helicity form factors using the light cone QCD sum rules. The uncertainties of the helicity form factors  come
from the borel parameter $M^{2}$, the charm quark mass $m_{c}$ and $\Phi_{\perp}$ twist-2 light cone distribution amplitude
of the axial vector meson. To extend the LCSR calculations to the full physical
region,  the extrapolated series expansions are used and  the low-lying $D$ meson resonances with ${1}^{+}$ and $1^{-}$
quantum numbers were utilized as the dominant poles. Based on the fitted form factors,
 predictions for the branching ratios of relevant semileptonic decays were reported and
 a comparison was
made between our results and other method estimations. Our calculation for branching ratio of
 $D^{+}\to K_{1}^{0}(1270)~e^{+}\nu_{e}$ decay is in
good agreement  with the BEPCII collider  measurement within errors at the mixing angle $\theta_{k}=-(36.68\pm6.30)^{\circ}$
\clearpage

\appendix
\begin{center}
{\Large \textbf{Appendix: Twist Function Definitions}}
\end{center}
In this appendix, we present the definitions for the two--parton LCDAs as well as the twist functions.
Two--particle chiral--even distribution amplitudes are given by
\cite{Kwei}:
\begin{eqnarray}\label{eq34}
\langle 0 | \bar{q}_{\alpha}(x)\,q_\delta (0)
| A(p, \varepsilon) \rangle &=& -\frac{i}{4}  \int_0^1 du~ e^{-i u  p'. x}\Bigg\{
f_{A} m_{A} \Bigg[ \not\! p\gamma_5
\frac{\varepsilon. x}{p.x} \Phi_\parallel(u) +\Bigg( \not\!
\varepsilon -\not\! p
\frac{\varepsilon. x}{p.x}\Bigg)\gamma_5 g_\perp^{(a)}(u) \nonumber\\
&-& \not\! x\gamma_5 \frac{\varepsilon. x}{2(p.x)^2} m_{A}^2
\phi_{b}(u) + \epsilon_{\mu\nu\rho\sigma} \varepsilon^{\nu}
p^{\rho} x^\sigma \gamma^\mu
\frac{g_\perp^{(v)}(u)}{4}\Bigg] \nonumber\\
&+& \,f^{\perp}_{A} \Bigg[ \frac{1}{2}( \not\! p\not\!\epsilon-
\not\!\epsilon \not\! p) \gamma_5\,\Phi_\perp(u) - \frac{1}{2}(
\not\! p\not\! x- \not\! x \not\! p ) \gamma_5 \frac{\epsilon.
x}{(p.x)^2} m_{A}^2 \bar
h_\parallel^{(t)} (u) \nonumber\\
&+& i \Big(\epsilon. x\Big) m_{A}^2 \gamma_5
\frac{h^{(p)}_\parallel (u)}{2} \Bigg]\Bigg\}_{\delta\alpha},
\end{eqnarray}
\begin{eqnarray}\label{eq36}
\langle 0|\bar{q}(x) \gamma_\mu \gamma_5
q'(0)|A(p, \varepsilon)\rangle &=& i f_{A} m_{A}\int_0^1 du \,  e^{-i u
p. x} \Bigg\{ p_\mu \frac{\varepsilon. x}{p. x}
\Phi_\parallel(u) +\left( \varepsilon_{\mu} -p_\mu
\frac{\varepsilon. x}{p.x}\right) g_\perp^{(a)}(u)
+{\cal O}(x^2) \Bigg\}~,\nonumber\\
\langle 0|\bar{q}(x) \gamma_\mu q'(0)|A (p, \varepsilon)\rangle
& = & - i f_{A}\, m_{A}~
\epsilon_{\mu\nu\rho\sigma} \varepsilon^{\nu} p^{\rho} x^\sigma
\int_0^1 du \, e^{-i u \, p. x}\Bigg\{
\frac{g_\perp^{(v)}(u)}{4}+{\cal O}(x^2)\Bigg\},
\end{eqnarray}
also, two--particle chiral--odd distribution amplitudes are defined
by:
\begin{eqnarray}\label{eq37}
\langle 0|\bar{q}(x) \sigma_{\mu\nu}\gamma_5
q'(0) |A (p, \varepsilon)\rangle & =&  f_{A}^{\perp} \int_0^1 du \, e^{-i u p'.
x} \Bigg\{(\varepsilon_{\mu} p_{\nu} - \varepsilon_{\nu}
p_{\mu}) \Phi_\perp(u) + \frac{{m^2_{A}}\,\varepsilon.
x}{(p. x)^2}(p_\mu x_\nu - p_\nu x_\mu) \bar{h}_\parallel^{(t)}
+{\cal O}(x^2)\Bigg\}, \nonumber \\
\langle 0|\bar{q}(x) \gamma_5 q'(0) |A(p, \varepsilon)\rangle
&=& f_{A}^\perp m_{A}^2 (\varepsilon. x)\int_0^1 du
\, e^{-i u p. x}\Bigg\{\frac{h_\parallel^{(p)}(u)}{2}+ {\cal
O}(x^2)\Bigg\}.
\end{eqnarray}
In these expressions, $f_{A}$ and $f_{A}^{\perp}$ are
decay constants of the axial vector meson $A$. We set
$f_{A}^{\perp}=f_{A}$ in $\mu=1~{\rm GeV}$, such that we
have
\begin{eqnarray}
\langle 0| \bar q(0) \sigma_{\mu\nu}\gamma_5
q'(0)   |A(p, \varepsilon)\rangle  =   a_0^{A}\,f_{A} \,
(\epsilon_{\mu} p_{\nu} - \epsilon_{\nu} p_{\mu}),
\end{eqnarray}
where $a_0^{\perp}$ refers to the zeroth Gegenbauer moments of
$\Phi_\perp$. It should be noted that $f_{A}$ is
scale--independent and conserves   $G$-parity, but
$f_{A}^{\perp}$ is scale--dependent and violates $G$-parity.

We take into account the approximate forms of twist-2 distributions
for the $A=a_1, K_{1A}$ states to be \cite{Kwei2}
\begin{eqnarray}
\Phi_\parallel(u) & = & 6 u \bar u \left[ 1 + 3 a_1^\parallel\, \xi +
a_2^\parallel\, \frac{3}{2} ( 5\xi^2  - 1 )
 \right], \label{eq:lcda-3p1-t2-1}\\
 \Phi_\perp(u) & = & 6 u \bar u \left[ a_0^\perp + 3 a_1^\perp\, \xi +
a_2^\perp\, \frac{3}{2} ( 5\xi^2  - 1 ) \right], \label{eq:lcda-3p1-t2-2}
\end{eqnarray}
and for the $A=b_1, K_{1B}$ to be
\begin{eqnarray}
 \Phi_\parallel(u) & = & 6 u \bar u \left[ a_0^\parallel + 3
a_1^\parallel\, \xi +
a_2^\parallel\, \frac{3}{2} ( 5\xi^2  - 1 ) \right], \label{eq:lcda-1p1-t2-1}\\
\Phi_\perp(u) & = & 6 u \bar u \left[ 1 + 3 a_1^\perp\, \xi +
a_2^\perp\, \frac{3}{2} ( 5\xi^2  - 1 ) \right],
\label{eq:lcda-1p1-t2-2}
\end{eqnarray}
where $\xi=2u-1$.

For the relevant two-parton twist-3 chiral-even LCDAs, we take the approximate
expressions up to conformal spin $9/2$ and ${\cal O}(m_s)$
\cite{Kwei2}:

\begin{eqnarray}
 g_\perp^{(a)}(u) & = &  \frac{3}{4}(1+\xi^2)
+ \frac{3}{2}\, a_1^\parallel\, \xi^3
 + \left(\frac{3}{7} \,
a_2^\parallel + 5 \zeta_{3, A}^V \right) \left(3\xi^2-1\right)
 \nonumber\\
& & {}+ \left( \frac{9}{112}\, a_2^\parallel + \frac{105}{16}\,
 \zeta_{3, A}^A - \frac{15}{64}\, \zeta_{3, A}^V \omega_{A}^V
 \right) \left( 35\xi^4 - 30 \xi^2 + 3\right) \nonumber\\
 & &
 + 5\Bigg[ \frac{21}{4}\zeta_{3, A}^V \sigma_{A}^V
  + \zeta_{3, A}^A \bigg(\lambda_{A}^A -\frac{3}{16}
 \sigma_{A}^A\Bigg) \Bigg]\xi(5\xi^2-3)
 \nonumber\\
& & {}-\frac{9}{2} \bar{a}_1^\perp
\,\widetilde{\delta}_+\,\left(\frac{3}{2}+\frac{3}{2}\xi^2+\ln u
 +\ln\bar{u}\right) - \frac{9}{2} \bar{a}_1^\perp\,\widetilde{\delta}_-\, (
3\xi + \ln\bar{u} - \ln u), \label{eq:ga-3p1}
\end{eqnarray}
\begin{eqnarray}
g_\perp^{(v)}(u) & = & 6 u \bar u \Bigg\{ 1 +
 \Bigg(a_1^\parallel + \frac{20}{3} \zeta_{3, A}^A
 \lambda_{A}^A\Bigg) \xi\nonumber\\
 && + \Bigg[\frac{1}{4}a_2^\parallel + \frac{5}{3}\,
 \zeta^V_{3, A} \left(1-\frac{3}{16}\, \omega^V_{A}\right)
 +\frac{35}{4} \zeta^A_{3, A}\Bigg] (5\xi^2-1) \nonumber\\
 &&+ \frac{35}{4}\Bigg(\zeta_{3, A}^V
 \sigma_{A}^V -\frac{1}{28}\zeta_{3, A}^A
 \sigma_{A}^A \Bigg) \xi(7\xi^2-3) \Bigg\}\nonumber\\
& & {} -18 \, \bar{a}_1^\perp\widetilde{\delta}_+ \,  (3u \bar{u} +
\bar{u} \ln \bar{u} + u \ln u ) - 18\,
\bar{a}_1^\perp\widetilde{\delta}_- \,  (u \bar{u}\xi + \bar{u} \ln \bar{u} -
u \ln u),
 \label{eq:gv-3p1}
 \end{eqnarray}
\begin{eqnarray}
h_\parallel^{(t)}(u) &= & 3a_0^\perp\xi^2+ \frac{3}{2}\,a_1^\perp
\,\xi (3 \xi^2-1) + \frac{3}{2} \Bigg[a_2^\perp \xi +
 \zeta^\perp_{3, A}\Bigg(5
-\frac{\omega_{A}^{\perp}}{2}\Bigg)\Bigg]\, \xi \,(5\xi^2-3)
\nonumber\\
 && +\frac{35}{4}\zeta^\perp_{3, A} \sigma^\perp_{A}
 (35\xi^4-30\xi^2+3)
  + 18 \bar{a}_2^\parallel
  \Bigg[\delta_+ \xi -\frac{5}{8}\delta_- (3\xi^2-1)\Bigg]\nonumber\\
 && -
  \frac{3}{2}\, \Bigg( \delta_+ \, \xi [2 +  \ln (\bar{u}u)]
   +\,\delta_- \, [ 1 + \xi \ln (\bar{u}/u) ]\Bigg)
   (1+ 6 \bar{a}_2^\parallel)
 ,\label{eq:ht-3p1}
 \end{eqnarray}
\begin{eqnarray}
h_\parallel^{(p)}(u) & = & 6u\bar u \Bigg\{ a_0^\perp +
\Bigg[a_1^\perp +5\zeta^\perp_{3, A}\Bigg(1-\frac{1}{40}(7\xi^2-3)
 \omega_{A}^{\perp} \Bigg)\Bigg] \xi \nonumber\\
 && \ \ \ \ \ \ \ + \Bigg( \frac{1}{4}a_2^\perp
 +
 \frac{35}{6} \zeta^\perp_{3, A} \sigma^\perp_{A} \Bigg)
 (5\xi^2-1)
 -5\bar{a}_2^\parallel
  \Bigg[\delta_+ \xi + \frac{3}{2} \delta_- (1-\bar{u} u) \Bigg]\Bigg\}
  \nonumber\\
 & & {}- 3[\, \delta_+\, (\bar{u} \ln \bar{u} - u \ln u)
 + \,\delta_-\,  ( u \bar{u} + \bar{u} \ln \bar{u} + u \ln u)]
 (1+ 6 \bar{a}_2^\parallel),
 \end{eqnarray}
for  $A=a_1, K_{1A}$ states, and
\begin{eqnarray}
 g_\perp^{(a)}(u) & = & \frac{3}{4} a_0^\parallel (1+\xi^2)
+ \frac{3}{2}\, a_1^\parallel\, \xi^3
 + 5\left[\frac{21}{4} \,\zeta_{3, A}^V
 + \zeta_{3, A}^A \Bigg(1-\frac{3}{16}\omega_{A}^A\Bigg)\right]
 \xi\left(5\xi^2-3\right)
 \nonumber\\
& & {}+ \frac{3}{16}\, a_2^\parallel \left(15\xi^4 -6 \xi^2 -1\right)
 + 5\, \zeta^V_{3, A}\lambda^V_{A}\left(3\xi^2 -1\right)
 \nonumber\\
& & {}+ \frac{105}{16}\left(\zeta^A_{3, A}\sigma^A_{A}
-\frac{1}{28} \zeta^V_{A}\sigma^V_{A}\right)
 \left(35\xi^4 -30 \xi^2 +3\right)\nonumber\\
 & & {}-15\bar{a}_2^\perp \bigg[ \widetilde{\delta}_+ \xi^3 +
 \frac{1}{2}\widetilde{\delta}_-(3\xi^2-1) \bigg] \nonumber\\
& & {}
  -\frac{3}{2}\,\bigg[\widetilde{\delta}_+\, ( 2 \xi + \ln\bar{u} -\ln u)
 +\, \widetilde{\delta}_-\,(2+\ln u + \ln\bar{u})\bigg](1+6\bar{a}_2^\perp)
 ,\label{eq:ga-1p1}
\end{eqnarray}
\begin{eqnarray}
g_\perp^{(v)}(u) & = & 6 u \bar u \Bigg\{ a_0^\parallel +
a_1^\parallel \xi +
 \Bigg[\frac{1}{4}a_2^\parallel
  +\frac{5}{3} \zeta^V_{3, A}
  \Bigg(\lambda^V_{A} -\frac{3}{16} \sigma^V_{A}\Bigg)
  +\frac{35}{4} \zeta^A_{3, A}\sigma^A_{A}\Bigg](5\xi^2-1) \nonumber\\
  & & {}  + \frac{20}{3}\,  \xi
 \left[\zeta^A_{3, A}
 + \frac{21}{16}
 \Bigg(\zeta^V_{3, A}- \frac{1}{28}\, \zeta^A_{3, A}\omega^A_{A}
  \Bigg)
 (7\xi^2-3)\right]\nonumber\\
 & & {} -5\, \bar{a}_2^\perp [2\widetilde\delta_+ \xi + \widetilde\delta_- (1+\xi^2)]
 \Bigg\}\nonumber\\
 & & {} - 6 \bigg[\, \widetilde{\delta}_+ \, (\bar{u} \ln\bar{u} -u\ln u )
  +\, \widetilde{\delta}_- \, (2u \bar{u} + \bar{u} \ln \bar{u} + u \ln u)\bigg]
  (1+6\bar{a}_2^\perp) ,
 \label{eq:gv-1p1}
\end{eqnarray}
\begin{eqnarray}
h_\parallel^{(t)}(u) &= & 3\xi^2+ \frac{3}{2}\,a_1^\perp \,\xi
(3\xi^2-1) + \Bigg[\frac{3}{2} a_2^\perp\, \xi
 + \frac{15}{2}\zeta^\perp_{3, A} \Bigg(\lambda^\perp_{A} - \frac{1}{10} \sigma^\perp_{A}\Bigg)
 \Bigg]
   \, \xi(5\xi^2-3) \nonumber\\
 && {} +\frac{35}{4}\zeta^\perp_{3, A}(35\xi^4-30\xi^2+3)\nonumber\\
 & & {} +\frac{9}{2} \bar{a}_1^\parallel\, \xi
 \Bigg[\delta_+\, (\ln u - \ln \bar{u} -3\xi)
  - \delta_-\,  \Bigg( \ln u + \ln \bar{u}
  +\frac{8}{3}\Bigg)\Bigg]\,,
\label{eq:ht-1p1}
\end{eqnarray}
\begin{eqnarray}
 h_\parallel^{(p)}(u) & = & 6u\bar u \Bigg\{ 1 + a_1^\perp \xi
 +  \left(\frac{1}{4}a_2^\perp +
 \frac{35}{6}\,\zeta^\perp_{3, A} \right)(5\xi^2-1)
  \nonumber\\
 & & {} +5\zeta^\perp_{3, A}
 \Bigg[\lambda^\perp_{A}-\frac{1}{40}(7\xi^3-3)\sigma^\perp_{A}
 \Bigg] \, \xi
 \Bigg\}  \nonumber\\
 & & {} -9\bar{a}_1^\parallel\, \delta_+\, (3 u \bar{u} + \bar{u} \ln
\bar{u} + u \ln u)
 -9\bar{a}_1^\parallel\,\delta_-\,  \Bigg( \frac{2}{3}\xi u\bar{u}
 + \bar{u} \ln \bar{u} - u \ln u \Bigg)\,, \label{eq:hp-1p1}
\end{eqnarray}
for  $A=b_1, K_{1B}$ states.
where
\begin{equation}
\widetilde{\delta}_\pm  ={f_{A}^{\perp}\over f_{A}}{m_{q_2} \pm m_{q_1}
\over m_{A}},\qquad \zeta_{3,A}^{V(A)} = \frac{f^{V(A)}_{3A}}{f_{A} m_{A}}.
\label{eq:parameters3}
\end{equation}

\end{document}